\documentclass[aps,prl,twocolumn,superscriptaddress,nofootinbib,preprintnumbers,10pt]{revtex4-1}

\pdfoutput=1
\usepackage[utf8]{inputenc}
\usepackage{graphicx,amsmath,amsfonts,amssymb,aas_macros,slashed}
\usepackage{bbold}
\usepackage{numprint}
\usepackage{enumitem}
 
\usepackage{graphicx,color,amsmath}
\usepackage{hyperref}

\allowdisplaybreaks

\makeatletter
\newcommand{\xRightarrow}[2][]{\ext@arrow 0359\Rightarrowfill@{#1}{#2}}
\makeatother

\DeclareMathOperator{\binomial}{binom}
\DeclareMathOperator{\gaussian}{gauss}

\newcommand{\cm}{\ensuremath{\mathrm{cm}}}

\newcommand{\eV}{\ensuremath{\mathrm{eV}}}
\newcommand{\keV}{\ensuremath{\mathrm{keV}}}
\newcommand{\MeV}{\ensuremath{\mathrm{MeV}}}
\newcommand{\GeV}{\ensuremath{\mathrm{GeV}}}

\newcommand{\me}[3]{\ensuremath \langle #1 | #2 | #3 \rangle   }

\usepackage{pgfplots}

\newcommand{\threej}[6]{\begin{pmatrix}#1&#3&#5\\#2&#4&#6\end{pmatrix}}

\usepackage{bm}
\usepackage{mathtools}

\renewcommand{\vec}[1]{\bm{#1}}

\usetikzlibrary{positioning,arrows,patterns}
\usetikzlibrary{decorations.markings}
\usetikzlibrary{calc}

\tikzset{
    photon/.style={decorate, decoration={snake}, draw=black},
    vector/.style={decorate, decoration={snake}, draw},
	provector/.style={decorate, decoration={snake,amplitude=2.5pt}, draw},
	antivector/.style={decorate, decoration={snake,amplitude=-2.5pt}, draw},
    fermion/.style={draw=black, postaction={decorate},
        decoration={markings,mark=at position .55 with {\arrow[draw=black]{>}}}},
    fermionbar/.style={draw=black, postaction={decorate},
        decoration={markings,mark=at position .55 with {\arrow[draw=black]{<}}}},
    fermionnoarrow/.style={draw=black},
    gluon/.style={decorate, draw=black,
        decoration={coil,amplitude=4pt, segment length=5pt}},
    scalar/.style={dashed,draw=black, postaction={decorate},
        decoration={markings,mark=at position .55 with {\arrow[draw=black]{>}}}},
    scalarbar/.style={dashed,draw=black, postaction={decorate},
        decoration={markings,mark=at position .55 with {\arrow[draw=black]{<}}}},
    scalartwo/.style={dotted,draw=black, postaction={decorate},
        decoration={markings,mark=at position .55 with {\arrow[draw=black]{>}}}},
    scalartwobar/.style={dotted,draw=black, postaction={decorate},
        decoration={markings,mark=at position .55 with {\arrow[draw=black]{<}}}},
    scalarnoarrow/.style={dashed,draw=black},
    electron/.style={draw=black, postaction={decorate},
        decoration={markings,mark=at position .55 with {\arrow[draw=black]{>}}}},
	bigvector/.style={decorate, decoration={snake,amplitude=4pt}, draw},
    vertex/.style={draw,shape=circle,fill=black,minimum size=1pt,inner sep=0pt},
    fermion2/.style={double, draw=black, postaction={decorate},
		decoration={markings,mark=at position .55 with {\arrow[draw=black]{>}}}},
    momentum/.style={draw=black,line width=0.15mm, postaction={decorate},
        decoration={markings,mark=at position 1 with {\arrow[draw=black]{>}}}}
}

\graphicspath{{figs/}}

\begin{document}

\title{On the relation between Migdal effect and dark matter-electron scattering \\ in isolated atoms and semiconductors}

\author{Rouven Essig}
\email{rouven.essig@stonybrook.edu}
\affiliation{C.N.~Yang Institute for Theoretical Physics, Stony Brook University, Stony Brook, NY 11794}

\author{Josef Pradler}
\email{josef.pradler@oeaw.ac.at}
\affiliation{Institute of High Energy Physics, Austrian Academy of Sciences, Nikolsdorfergasse 18, 1050 Vienna, Austria}

\author{Mukul Sholapurkar}
\email{mukul.sholapurkar@stonybrook.edu}
\affiliation{C.N.~Yang Institute for Theoretical Physics, Stony Brook University, Stony Brook, NY 11794}

\author{Tien-Tien Yu}
\email{tientien@uoregon.edu}
\affiliation{Department of Physics,
University of Oregon, Eugene, Oregon 97403}

\preprint{YITP-19-23}

\begin{abstract}
%
  A key strategy for the direct detection of sub-GeV dark matter is to search for small ionization signals. These can arise from dark matter-electron scattering or when the dark matter-nucleus scattering process is accompanied by a ``Migdal'' electron.  We show that the theoretical descriptions of both processes are closely related, which allows for a principal mapping between dark matter-electron and dark matter-nucleus scattering rates once the dark matter interactions with matter are specified.  We explore this parametric relationship for noble-liquid targets and, for the first time, provide an estimate of the ``Migdal'' ionization rate in semiconductors that is based on evaluating a crystal form factor that accounts for the semiconductor band structure.  We also present new dark-matter-nucleus scattering limits down to dark matter masses of 500~keV using published data from XENON10, XENON100, and a SENSEI prototype Skipper-CCD. For a dark photon mediator, the dark matter-electron scattering rates dominate over the Migdal rates for dark matter masses below 100~MeV. We also provide projections for proposed experiments with xenon and silicon targets. 
\end{abstract}

\maketitle

\paragraph{\textbf{Introduction.}}

Detectors searching for direct signals from dark matter (DM) are conventionally optimized to look for electroweak-scale DM that scatters elastically off atomic nuclei.  
The detectors typically lose sensitivity rapidly for DM masses below a few GeV, due to an inefficient energy transfer from the DM to the recoiling nucleus.  
However, the kinematic limitations are lifted when the DM-nucleus scattering process is accompanied by the irreducible simultaneous emission of a ``bremsstrahlung" photon~\cite{Kouvaris:2016afs} or a ``Migdal"-electron~\cite{Ibe:2017yqa},  
or by considering alternative interactions such as DM-electron scattering~\cite{Essig:2011nj}. 
In all cases the entire energy of relative motion between the atom and DM can in principle be transferred to the outgoing photon or electron. 
These signals have already opened a new pathway for current detectors to register DM scattering on nuclei~\cite{Akerib:2018hck,Armengaud:2019kfj,Liu:2019kzq,Aprile:2019jmx} for DM masses below 100~MeV (driven primarily by the stronger Migdal effect), and DM scattering on electrons~\cite{Essig:2012yx,Essig:2017kqs,Tiffenberg:2017aac,Romani:2017iwi,Crisler:2018gci, Agnese:2018col, Agnes:2018oej,  Abramoff:2019dfb, Aguilar-Arevalo:2019wdi} for masses as low as 500~keV.  The sensitivity to sub-GeV DM is expected to improve significantly over the next few years as new detectors with an ultralow ionization threshold are being developed~\cite{Tiffenberg:2017aac,Settimo:2018qcm}. For related and distinct direct-detection ideas to probe sub-GeV DM, see e.g.~\cite{Vergados:2004bm,Moustakidis:2005gx,Ejiri:2005aj,Bernabei:2007jz,Essig:2015cda,Graham:2012su,An:2014twa,Aprile:2014eoa,Lee:2015qva,Hochberg:2015pha,Hochberg:2015fth,Aguilar-Arevalo:2016zop,Bloch:2016sjj,Cavoto:2016lqo,Derenzo:2016fse,Essig:2016crl,Hochberg:2016ntt,Hochberg:2016ajh,Hochberg:2016sqx,Budnik:2017sbu,Bunting:2017net,Cavoto:2017otc,Fichet:2017bng,Knapen:2017ekk,Hochberg:2017wce,Dolan:2017xbu,Bringmann:2018cvk,Ema:2018bih,Emken:2019tni,Bell:2019egg,Cappiello:2019qsw}. 

The theoretical description of the bremsstrahlung and Migdal effect is so far exclusively tied to a picture where DM scatters on a single, isolated atom~\cite{Kouvaris:2016afs,Ibe:2017yqa},  
which for inner-shell electrons should provide a correct estimate of the expected signal rate.  
For the outer-shell electrons, one might have some hope that the isolated atom calculations are still valid for noble-liquids, but the complicated electronic band structure for semiconductor targets definitely calls for a different approach. The results presented for noble liquids in~\cite{Akerib:2018hck,Aprile:2019jmx} and for germanium in~\cite{Armengaud:2019kfj,Liu:2019kzq} restricted themselves to inner-shell electrons and had higher detector threshold than those needed to see a dominant signal from the outer shells. 
However, several experiments already have sensitivity to one or a few electrons, and a complete understanding of the Migdal and bremsstrahlung effect will maximize our sensitivity to nuclear scattering. 
On the other hand, DM-electron scattering has already been calculated with a detailed accounting of the band structure in~\cite{Essig:2015cda}. 

In this work, our first goal is to show that the theoretical calculation of the rates for the Migdal effect and for DM-electron scattering are closely related.  This observation allows us to take the first steps towards calculating the Migdal effect in semiconductors, thereby overcoming the previous limitations in the theoretical description.  It allows us also to calculate the first DM-nucleus scattering limits down to DM masses of 500~keV using published SENSEI results with a single-electron threshold~\cite{Crisler:2018gci}.  A second goal is to use the observed single- and few-electron events from XENON10 to calculate the Migdal effect, which provides a DM-nuclear scattering limit down to 5~MeV, lower in mass than other published constraints.  We compare this limit to XENON100 and XENON1T. Finally, a third goal is to contrast the constraints from DM-electron scattering with those from the Migdal effect in DM-nuclear scattering when the DM interations with ordinary matter are mediated by a dark photon, which gives rise to both signals simultaneously. 
A more detailed exposition will be provided in a companion paper~\cite{MigdalPRD}.

To appreciate the connection between the DM-electron and Migdal processes, consider the nucleus receiving 
a sudden three momentum transfer $\vec q$ in the scattering with DM.
The probability for promptly ejecting an electron from an atom that was initially at rest 
in state $|i\rangle$ is then found by projecting the boosted electron cloud onto the desired final state~$|f\rangle$,
\begin{align}
  \label{eq:boost}
  \me{f}{e^{i \frac{m_e}{m_N} \vec q  \cdot \sum_\alpha \vec x^{(\alpha)}} }{i}
  \simeq    \frac{i}{e} \frac{m_e}{ m_N} \vec q \cdot \vec d_{fi} \quad (i\neq f).
\end{align}
Here, $m_e$ ($m_N$) is the mass of the electron (nucleus) and the sum
runs over all electron positions $\vec x^{(\alpha)}$. The
exponential on the left hand side is indeed the electron boost operator as
$\vec v_N' = \vec q / m_N$ is the velocity of the recoiling
nucleus. On the right hand side, we have used the dipole approximation,
noting that $\frac{m_e}{m_N} \vec q \cdot \vec x^{(\alpha)} \ll 1$ in all
cases of interest;
$\vec d_{fi} = \me{f}{e \sum_{\alpha} \vec x^{(\alpha)}}{i} $ is the
atom transition dipole moment.
In single-electron transitions only one term will contribute in the latter sum, and 
$\vec d_{fi} $ is to be replaced  by $ \vec d_{fi} \to \vec d^{(\beta)}_{fi} = \me{f}{e \vec  x^{(\beta)}}{i}$, where $\vec x^{(\beta)}$ is the coordinate of the electron undergoing the transition.

In turn, in the calculation of the ionization probability
from the {\it direct} interaction of DM with an electron at coordinate
$\vec x^{(\beta)}$, the following matrix element enters,
\begin{align}
  \label{eq:direct-kick}
  \me{f}{e^{i  \vec q  \cdot  \vec x^{(\beta)}} }{i}
  \simeq   \frac{i}{e} \vec q \cdot \vec d^{(\beta)}_{fi} \quad (i\neq f).
\end{align}
Here the momentum transfer $\vec q$, \textit{i.e.}~the momentum lost
by the DM particle, is directly picked up by the electron. Comparing
(\ref{eq:boost}) with (\ref{eq:direct-kick}) clearly exposes the
similarity of Migdal-effect and DM-electron scattering. However,  at the same time 
this also shows that the atom is probed at vastly different momentum transfers in the respective processes, a
factor of $\sim 10^{-3}/A$ softer in the Migdal case.

\paragraph{\textbf{Migdal effect in isolated atoms.}}
We now establish the precise relation
between DM-electron scattering and Migdal effect for isolated atoms. 
In an isolated atom, the bound initial-state electron wave functions are specified by their principal and orbital quantum numbers $n$ and $l$. We are interested in transitions to continuum final-states that are characterized by momentum $p_e=\sqrt{2m_e E_e}$ and orbital quantum number $l'$. We define the ionization probability for such a scenario as
\begin{equation}
\left|  \me{p_e,l'}{e^{i \frac{m_e}{m_N} \vec q  \cdot \sum_\alpha \vec x^{(\alpha)}} }{n,l}\right|^2=\frac{1}{2\pi}\frac{dp_{nl\to E_e}(q)}{dE_e}\, .
\end{equation}
A key observation of this letter is that the electron-ionization probability ${dp_{nl\to E_e}(q)}/{dE_e}$ is related to a dimensionless form factor  $ |f^{\rm ion}_{nl}(p_e, q_e)|^2$ that was previously defined in the context of DM-electron scattering~\cite{Essig:2011nj,Essig:2012yx}. Concretely, we find
\begin{align}
  \label{eq:ionizationprobFion2}
  \frac{dp_{nl\to E_{e}}}{d\ln E_{e}}  = \frac{\pi}{2} |f^{\rm ion}_{nl}(p_e, q_e)|^2  \,,
\end{align}
where the form factor is evaluated at $q_e \simeq \frac{m_e}{m_N} q$.
Capturing the Migdal process hence becomes a question of calculating
the form factor accurately. It is given by~\cite{Essig:2011nj,Essig:2012yx} 
\begin{align}
  \label{eq:fion}
  |f^{\rm ion}_{nl}(p_e,q_e)|^2 & = \frac{4 p_e}{2\pi}
                                  \sum_{L=1}^{\infty} \sum_{l'=0}^{\infty} \frac{m_l}{2} (2l'+1)(2L+1) \nonumber \\
                                &\times \threej{l}{0}{l'}{0}{L}{0}^2  \left| \int dr\, r^2 R_{p_el'} R_{nl} j_L(q_e r) \right|^2 ,
\end{align}
where $R_{nl}$ is the radial wavefunction of the $n,l$ bound state 
orbital with shell occupancy $m_l\leq 2(2l+1)$, $R_{p_el'}$ is the continuum wave function of the electron
with momentum $p_e$, and $j_L$ is the spherical Bessel function of the
first kind.%
\footnote{The normalization is
  $\int dr\, r^2 R_{n'l'}R_{nl} = \delta_{nn'}\delta_{ll'} $ for bound
  states and
  $\int dr\, r^2 R_{kl} R_{k'l'} = (2\pi) \delta(k-k')\delta_{ll'}$
   for states in the continuum.}  
The $L=0$ term does not induce a transition and is
omitted in sum for clarity.

A detailed calculation of $dp_{n,l\to E_e}/dE_e$ was presented in
\cite{Ibe:2017yqa} using a fully relativistic formulation. Numerical results were calculated in the dipole
approximation, which in the above formulation would amount to
expanding $j_1(q_er)\simeq q_er/3$ and taking the $L=1$
term. In~\cite{Ibe:2017yqa}, bound and continuum state wave functions
of angular momentum $j = l \pm s$ were generated with the FAC atomic
code~\cite{doi:10.1139/p07-197}. Following this approach, we find
agreement with their results. Since electron energies remain in the non-relativistic domain (with small corrections for $n=1$ states in elements with large atomic number), a formulation where differences from
spin are unresolved will capture the process to sufficient accuracy.
Such approximation has been chosen in Eq.~(\ref{eq:fion}) where $l$ labels
two spin-degenerate states.%
\footnote{When the momentum transfer exceeds several MeV, as would be the case 
for DM with mass above the GeV-scale, the non-relativistic approximation
underestimates the form-factor~\cite{Roberts:2016xfw} and a
full treatment becomes necessary.}

\begin{figure}[tb]
\centering
\includegraphics[width=\columnwidth]{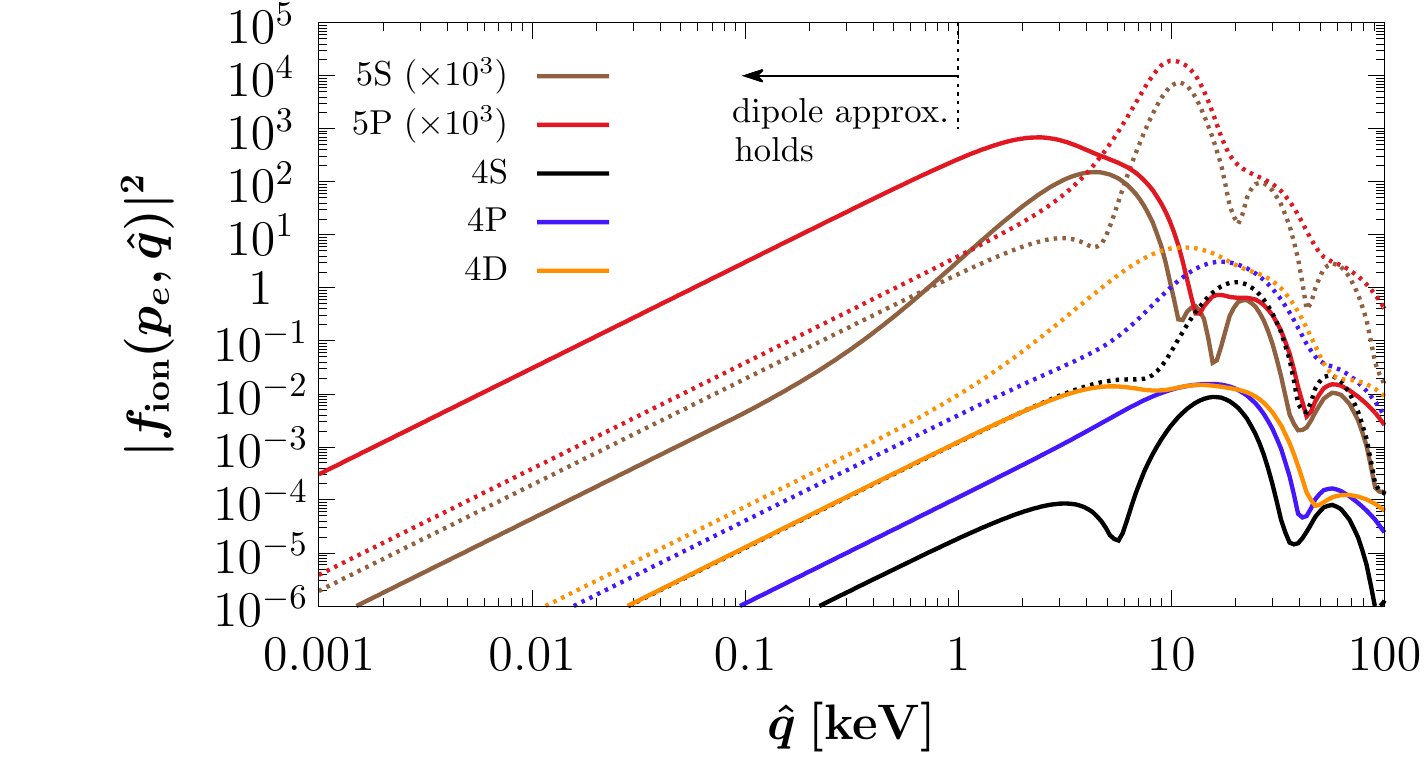} 
\caption{Ionization form factors for the $n=4$ and $n=5$shells in xenon as a function of 
 $q$ for  $E_e=1\,\eV$ (100\,\eV) shown by the solid (dotted) lines. 
Below $q=1\,\keV$, the form factor scales as $q^2$ and the dipole approximation in Eq.~(\ref{eq:dipole-rescaling}) is valid.
At higher momentum transfers, $ |f^{\rm ion}_{nl}(p_e,q)|^2 $ tends to  peak before becoming strongly suppressed; $\hat q=q_e$ for Migdal whereas $\hat q=q$ for DM-electron scattering.}
\label{fig:fion}
\end{figure}

The dipole approximation in Eq.~(\ref{eq:fion}) results in the following scaling-relation in the form-factor
\begin{align}
\label{eq:dipole-rescaling}
     |f^{\rm ion}_{nl}(p_e,q)|^2 = \frac{q^2}{q_0^2} \times   |f^{\rm ion}_{nl}(p_e,q_0)|^2 \quad (\text{dipole approx.})\, .
\end{align}
Here, $q_0$ is a reference momentum at which the scaling-relation holds, {\it i.e.}, $q_0 r_e\ll 1$, where $r_e$ is the electron radius; for xenon atoms, $q_0\lesssim 1\,\keV$.
The scaling-relation breaks down once $qr_e\gtrsim 1$ as
higher multipole ($L>1$) contributions in Eq.~(\ref{eq:fion}) become important and must be included; in our numerical results we include terms up to $L=15$ using
FAC-generated continuum and bound-state wave functions. 
The general behavior of $ |f^{\rm ion}_{nl}(p_e,q)|^2 $ is that it peaks at $q\sim \keV$ before becoming strongly suppressed in the high-$q$ region of tens of keV and above. This behavior is illustrated in Fig.~\ref{fig:fion} where the area of validity
of Eq.~(\ref{eq:dipole-rescaling}) is indicated.

\paragraph{\textbf{Cross sections.}}

\begin{figure*}\centering
\includegraphics[width=\textwidth, angle=0]{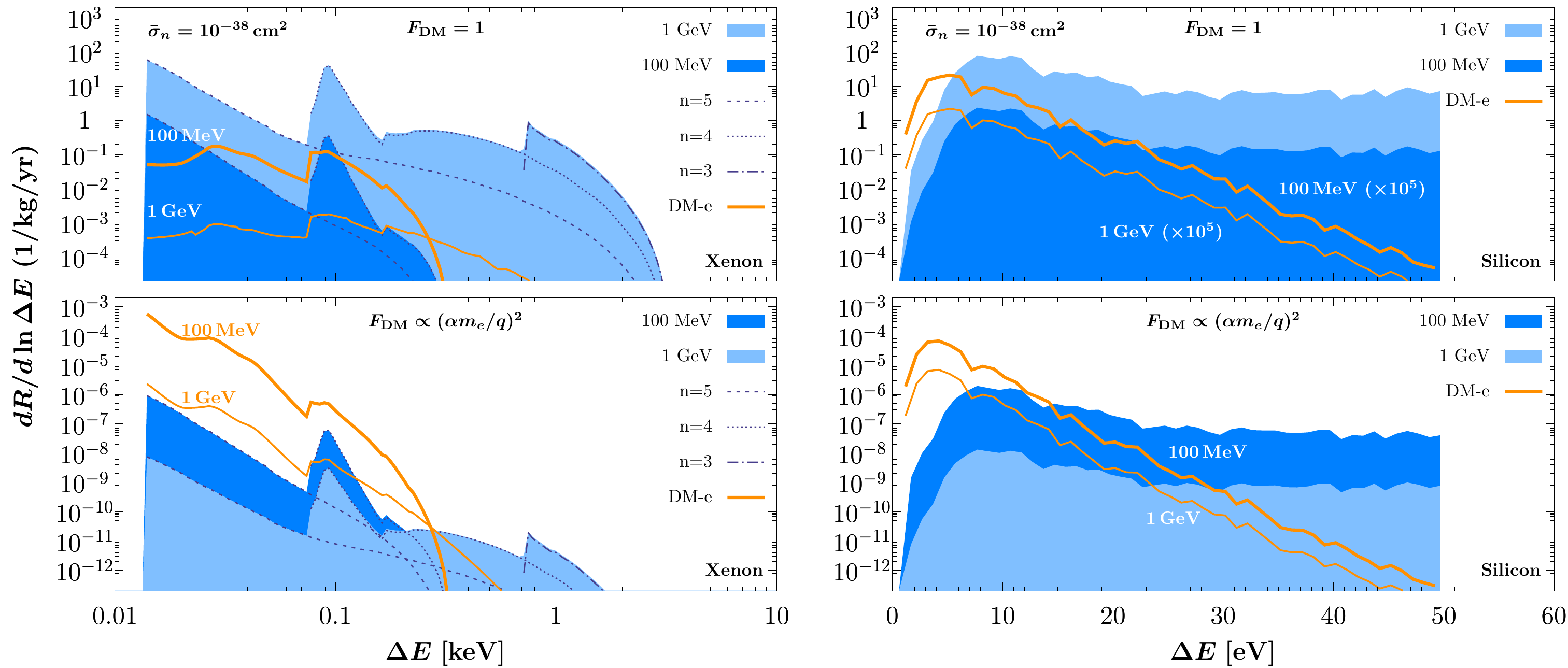}%
\caption{\small {\it Left:} Light (dark) shaded regions show the cumulative Migdal event rates in xenon for $m_\chi = 1\,\GeV$ ($100\,\MeV$) as a function of deposited electronic energy $\Delta E$ for  contact interactions (top panel) and light mediator scaling (bottom panel).  Individual contributions  from the various shells are indicated by the various dashed lines. For the light mediator a reference momentum of $q_0=\alpha m_e$ and $m_V =0$ is chosen. The rates for DM-electron scattering for  $m_\chi = 1\,\GeV$ ($100\,\MeV$) are shown by the thick (thin) orange solid lines. {\it Right:} Same as left panel, but for silicon and evaluated using our formulation of the Migdal effect in semiconductors. The rates are averaged over a 1~eV energy bin.}
\label{sigavg}
\end{figure*}

The prompt ionization resulting from DM-nuclear scattering with
nuclear recoil energy $E_R =  q^2/2m_N$ leads to a double-differential cross section in $E_R$ and $E_{e}$. Away from kinematic
endpoints, the cross-section can be factorized as the elastic cross-section times the electron ionization probability~\cite{Ibe:2017yqa},
\begin{align}
  \label{eq:double-differential}
  \frac{d\sigma_{n,l}}{dE_R dE_{e}} \simeq \frac{d\sigma}{dE_R}
  \times  \frac{1}{2\pi} \frac{dp_{n,l\to E_e}}{dE_e}\,. 
\end{align}
The elastic DM-nucleus recoil cross section 
is given by
\begin{align}
 \frac{d\sigma}{dE_R} = \frac{\bar\sigma_n m_N}{2\mu_n^2v^2}   [f_p Z + f_n (A-Z)]^2 |F_{\rm DM}(q)|^2 |F_N(q)|^2  \,,
\end{align}
where $Z$ is the nuclear charge, $f_{p,n}$ are dimensionless DM
couplings to the proton (neutron), and $\mu_n$ is the DM-nucleon reduced
mass. The DM-nucleon cross section is defined as
$  \bar \sigma_n \equiv \mu_n^2 
\overline{ |\mathcal{M}_n(q= q_0)|^2}  / (16\pi m_{\chi}^2 m_n^2 )  $,
where $q_0=\alpha m_e$ is a reference momentum scale.
The model dependence in the DM-interaction with the nucleus is
absorbed into the definition of a form factor,
\begin{align}
  |F_{\rm DM}(q)|^2\equiv
  \frac{ \overline{ |\mathcal{M}_n(q)|^2} } { \overline{ |\mathcal{M}_n(q= q_0)|^2}} =
  \begin{cases}
    1 & \text{contact} \\
    \frac{m_V^2 + q_0^2}{m_V^2 + q^2} & \text{light med.}  \end{cases}\,,
\end{align}
where $ \overline{ |\mathcal{M}_n(q)|^2}$ is the matrix element for
scattering on a free nucleon. In the second equality, the two given
examples correspond to contact interactions and to the exchange of a
 vector mediator with mass $m_V$ and a reference momentum $ q_0$.
For sub-GeV DM, the momentum
transfer does not resolve the nucleus, $q R_N\ll 1$, and the nuclear
form factor $|F_N(q)|^2 = 1$ to good
approximation.

Integrating Eq.~(\ref{eq:double-differential}) over $E_R$ then yields
the Migdal electron spectrum, where the kinematic limits of $E_R$ are fixed by energy conservation. 
The cross section is then averaged over the galactic
velocity distribution of DM, boosted into the laboratory frame of the
detector, $ g_{\rm det}(\vec v) $~\cite{Lewin:1995rx},
$ { d \langle\sigma_{n,l} v \rangle }/{ d E_{e} } = \int_{|\vec
  v|>v_{\rm min}} d^3\vec v\, g_{\rm det}(\vec v) { d \sigma_{n,l} v
}/{ d E_{e} } $. Here, $v_{\min}$ is the minimum velocity such that
the DM energy is larger than the total energy received by the electron,
$v_{\min} = \sqrt{2\Delta E_{n,l}/\mu_N}$; and $\Delta E_{n,l} = E_e + |E_{n,l}| $, where $E_{n,l} $ is the binding energy, which we take from~\cite{Essig:2017kqs}.
Finally, the event rate is given by 
$dR_{n,l}/dE_e = N_T (\rho_{\rm dm}/m_\chi)  { d \langle\sigma_{n,l} v \rangle }/{ d E_{e} } $ 
where $N_T$ is the number of targets per kg of detector material and $\rho_{\rm dm } \simeq 0.3~\GeV/\cm^3$ is the local DM density. 
The resulting rate as a function of $\Delta E_{n,l}$, summed over all $l$ for the various $n$-shells of xenon, is shown in Fig.~\ref{sigavg} (left) for contact interactions (top panel) and light mediator exchange (bottom panel). 

The velocity-averaged ionization cross-section can be written in a form analogous to the one used in DM-electron scattering~\cite{Essig:2011nj} by 
exchanging the integration order
between $d^3\vec v$ and $dE_R$ ($=q dq/m_N$), 
\begin{align}
    \label{eq:ionizationcsDMeversion}
  \frac{  d \langle\sigma_{n,l} v \rangle }{ d\ln E_{e}  }  &
  =\frac{\bar\sigma_n}{8\mu_n^2} [f_pZ + f_n (A-Z)]^2 \int dq\, \left[ q |F_N(q)|^2      \right. \nonumber \\ &
 \!\!\!\!\!\!\!\!\!\!   \left.  \times |F_{\rm DM}(q)|^2  |f^{\rm ion}_{nl}(p_e, q_e)|^2 \eta(v_{\rm min}(q,\Delta E_{n,l})) \right]\,,
\end{align}
where $\eta(v_{\rm min})$ is the usual velocity average of the inverse
speed,
$\eta(v_{\rm min}) = \langle \frac{1}{v} \Theta(v-v_{\rm min})
\rangle_{g_{\rm det}} $, but now the minimum velocity becomes a
function of both $q$ and $\Delta E_{n,l}$,
  $v_{\rm min}(q,\Delta E_{n,l}) \simeq q/(2m_\chi) + \Delta E_{n,l}/{q} $,
  where we
  have made the approximation $\mu_N \simeq m_\chi$ in the expression for 
  $v_{\rm min}$ as is applicable for sub-GeV DM.
Comparing Eq.~(\ref{eq:ionizationcsDMeversion}) to the cross-section for DM-electron scattering~\cite{Essig:2011nj},
\begin{align}
    \label{eq:DMeCS}
  \frac{  d \langle\sigma^{\rm DM-e}_{n,l} v \rangle }{ d\ln E_{e}  }  & =  \frac{\bar\sigma_e}{8\mu_e^2} \int dq\, \left[ q  |F_{\rm DM}(q)|^2
              |f^{\rm ion}_{nl}(p_e, q)|^2                                                         \right.
  \nonumber   \\ & 
                   \left.    \qquad \qquad  \quad \times
                   \eta(v_{\rm min}(q,\Delta E_{n,l})) \right] \,,
\end{align}
exposes the striking similarity of both processes, with a key difference. The DM form factor is now evaluated at the scale $q$ not
$q_e$ as the electron directly picks up the entire momentum transfer.

Any direct comparison between the two processes is model-dependent as
it requires specifying both $\bar\sigma_e$ and $\bar\sigma_n$. Consider for
concreteness the scattering mediated by a dark photon, coupling to
electric charge $e$ with a strength $\varepsilon e$. Then,
\begin{align}\label{eq:darkphoton}
  \bar \sigma_e = \frac{16\pi \varepsilon^2 \alpha \alpha_D \mu_e^2}{(q_0^2 + m_V^2)^2}, \quad \bar\sigma_p = \frac{\mu_p^2 }{\mu_e^2} \bar \sigma_e , 
\end{align}
where $\bar\sigma_p$ is the DM-proton scattering cross section
$(f_p= 1, f_n=0)$, $\alpha_D = g_V^2/4\pi$, where $g_V$ is the DM-dark
photon coupling, and for consistency we define $\bar\sigma_n = (Z^2/A^2)\bar\sigma_p$.
In the expression for the Migdal effect, $|F_N(q)|^2$ needs to account for the screening of the nuclear
electric potential, see \textit{e.g.}~\cite{Emken:2019tni}.  It is, however, of little relevance for the typically considered momentum transfers, and we will neglect it to exhibit better the parametric relations. 

For a dark photon mediator, the ratio of differential cross sections in Eqs.~(\ref{eq:ionizationcsDMeversion}) and~(\ref{eq:DMeCS}) becomes 
\begin{align}\label{eq:ratio}
   \frac{d\langle \sigma_{n,l}^{\rm{Migdal}}v\rangle/(d \ln{E_e} dq)}{d\langle \sigma_{n,l}^{\rm{DM-e}}v\rangle/(d \ln{E_e} dq)} =  Z^2 \times  \frac{|f^{\rm ion}_{nl}(p_e, q_e)|^2}{|f^{\rm ion}_{nl}(p_e, q)|^2}.
\end{align}
At this point, one may be tempted to exploit the dipole scaling relationship in Eq.~(\ref{eq:dipole-rescaling}) to conclude that the RHS scales as 
$Z^2 m_e^2/m_N^2 \sim 10^{-6} (Z/A)^2$ implying that the rate for Migdal ionization is always much smaller than for DM-electron scattering. 
However, whereas the dipole approximation  is always valid for the Migdal effect, since $q_e r_e \ll 1$, kinematic arguments imply that for DM-electron scattering $\Delta E_{n,l}/v_{\rm{max}}  \lesssim q \lesssim 2 \mu_N v_{\rm{max}}$.  In this region, $q \gtrsim 1/r_e$ and the dipole scaling breaks down. Inspection of Fig.~\ref{fig:fion} shows that at large enough momentum transfer the form factors become strongly suppressed, so that at some critical value  $q^{\rm{crit}} > 1/r_e$, $|f^{\rm ion}_{nl}(p_e, q^{\rm{crit}})|^2 = |f^{\rm ion}_{nl}(p_e, q_e^{\rm{crit}})|^2 \times Z^2 $ (with $q_e^{\rm{crit}}=\frac{m_e}{m_N}q^{\rm{crit}}$) is certainly met. 
Hence, for the differential rates in Eq.~(\ref{eq:ratio}), DM-electron scattering only dominates over Migdal scattering for $q < q^{\rm{crit}}$, while the Migdal effect dominates over DM-electron scattering for higher mass DM. Clearly, both effects need to be taken into account to derive accurate DM constraints.  
For contact interactions, the Migdal effect dominates for $m_{\chi} \gtrsim (q^{\rm{crit}})^2 /(2 \Delta E_{n,l})$. For long-range interactions, DM-electron interactions dominate essentially for all masses as there is a bias towards lower momenta introduced by $|F_{\rm DM}(q)|^2$. 

\begin{figure}[tb]
\includegraphics[width=\columnwidth, angle=0]{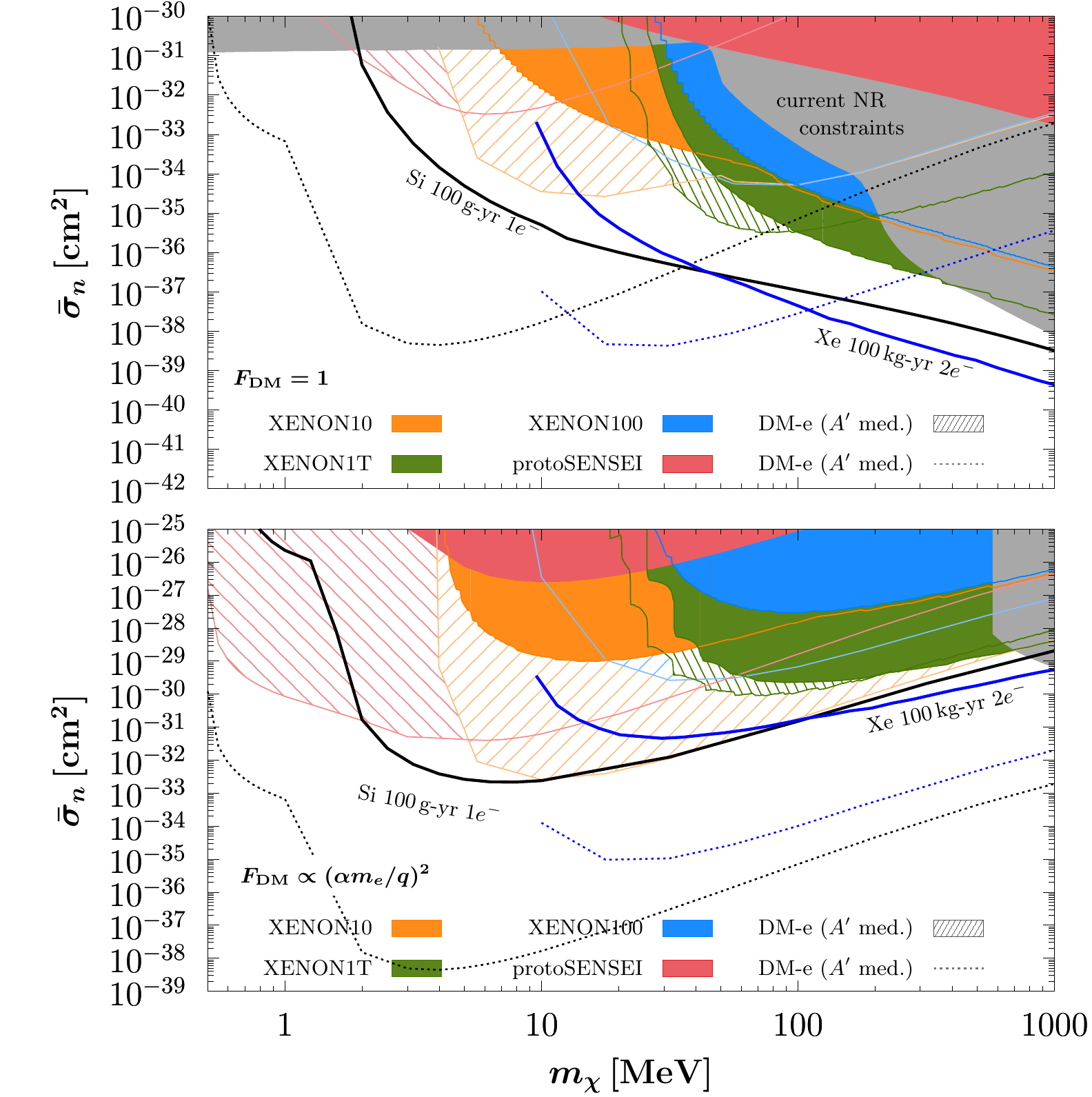}%
\caption{\small The 90\%CL limits with $f_n=f_p=1$ for a heavy (top panel) and light (bottom panel) mediator  from XENON10, XENON100, and XENON1T as well as projections for a 100~g-year silicon detector and LBECA, a 100~kg-year xenon detector, with 2-electron thresholds. The solid regions correspond to Migdal-scattering, while the hatched regions are the translation of DM-electron scattering constraints in a dark photon model. In the top panel, the gray-shaded region corresponds to current constraints on $\overline\sigma_n$ from CRESST III~\cite{Abdelhameed:2019hmk}, organic liquid scintillator~\cite{Collar:2018ydf}, DarkSide-50~\cite{Agnes:2018ves}, CDEX~\cite{Liu:2019kzq}, XENON1T~\cite{Aprile:2018dbl,Aprile:2019xxb}, and a recast of XENON1T data for cosmic-ray up-scatter~\cite{Bringmann:2018cvk}. In the bottom panel, the gray-shaded region corresponds to constraints from LUX~\cite{Akerib:2018hck} and Panda-X~\cite{Ren:2018gyx}.}
\label{fig:constraints}
\end{figure}

\paragraph{\textbf{Application to semiconductors.}}

Until now, all studies of the Migdal effect in semiconductor targets considered isolated atoms. 
Using our above results and working along the lines of~\cite{Essig:2015cda}  that previously exposed the 
connection for DM-electron scattering between isolated atoms and semiconductors, we are in a position to arrive at a rate equation for Migdal scattering in semiconductors.
The replacement to be made in Eq.~(\ref{eq:ionizationcsDMeversion}) to obtain  the analogous cross section $d \langle\sigma_{\rm crystal} v \rangle/d\ln E_{e}$ is
\begin{equation}
\label{eq:SCreplacement}
    |f_{n,l}^{\rm ion}(q_e,E_e)|^2 \to  \frac{8\alpha m_e^2 E_e}{q_e^3}\times |f_{\rm crystal}(q_e,E_e)|^2\, ,
\end{equation}
where $|f_{\text{crystal}}(q_e,E_e)|^2 $ is a dimensionless crystal form factor obtained in~\cite{Essig:2015cda,QEdark}. The ionization 
rate is then given by 
$ d R_{\text{crystal}}/d \ln E_e = N_{\text{cell}} (\rho_{\chi}/m_{\chi}) d \langle\sigma_{\rm crystal} v \rangle /{ d\ln E_{e}  }   $ where  $N_{\rm cell}\equiv M_{\rm target}/M_{\rm cell}=N_T/2$  is the number of unit cells in the crystal. 

In deriving Eq.~(\ref{eq:SCreplacement}) we made use of the dipole scaling relation~(\ref{eq:dipole-rescaling}).  This scaling is only approximate: due to the crystal's electronic structure, there are only discrete values of $q_e$ available for a given $E_e$ . 
Moreover, because of small $q_e$, the Migdal effect favors direct-gap transitions ($\gtrsim$3~eV) over indirect-gap transition ($\gtrsim$1.2~eV), affecting the threshold behavior with respect to DM-e scattering. A more refined computation of the crystal form-factor at low momentum is warranted and left to~\cite{MigdalPRD}. Here we show a proof of concept of the mapping between the ionization and crystal form-factors, and calculate the first projections on $\overline\sigma_n$ from the Migdal effect in silicon. 
 Since the dipole approximation is valid when $q r_e\ll 1$, where $r_e\sim 1/(\alpha m_e)$ in silicon, we anchor our form-factors at $q_0=0.5 \alpha m_e$ and take the average value of $f_{\rm ion}$ in a neighborhood around $q_0$. The {\tt QEdark}~\cite{QEdark} form-factors are calculated up to $E_e=50$ eV, which is sufficient for DM-electron scattering. However, from Fig.~\ref{sigavg} (right), we see that Migdal-scattering extends out to higher values of $E_e$ and the truncation at $E_e=50$ eV underestimates the rate; in principle, one can combine the form-factors calculated with relativistic FAC wavefunctions, which are sufficiently accurate for $E_e\gtrsim 50$ eV, to obtain a more complete spectrum~\cite{MigdalPRD}. Here, we show the constraints using the {\tt QEdark} form factors only. 

\paragraph{\textbf{Constraints and Projections.}}

We now apply the above results to derive new constraints and projections from direct detection experiments.  
For xenon detectors, the ionization (S2) signal is obtained by folding $dR/dE_e$ with the
probability to produce $S2$ photo-electrons (PE) given a
deposited energy $\Delta E_{nl}$ and a detection efficiency 
$\epsilon(S2)$, 
$dR_{nl}/dS2 = \epsilon(S2) \int dE_e\, P(S2|\Delta E_{nl})  dR_{nl}/dE_e $. 
Here, 
$   P(S2|  \Delta E_{nl}  ) = \sum_{n_e^{\rm surv},n_e}
  P(S2|n_e^{\rm surv})   P(n_e^{\rm surv} | n_e) 
  P(n_e | \langle n_e \rangle )$.  
The number of electrons escaping the interaction point $n_e$ are assumed to 
follow a binomial distribution, $ P(n_e | \langle n_e \rangle )  = \binomial (n_e |  N_Q , f_e ) $ with $N_Q = \Delta E_{nl} /(13.8~\eV)$~\cite{Dahl:2009nta,Akerib:2016qlr} trials and a 
single event probability $f_e = \langle n_e \rangle /N_Q$; the mean number of 
electrons $\langle n_e \rangle $ is either modeled following~\cite{Essig:2012yx} or, for  $\Delta E_{nl} > 0.19\,\keV$, obtained from the measured charge yield in~\cite{Akerib:2017hph}. For $ P(n_e^{\rm surv} | n_e)  $, we assume that 80\% (100\%) of electrons survive the drift in XENON1T/XENON100 (XENON10). Those electrons induce 
scintillation at the liquid-gas interface, described by a Gaussian, $  P(S2|n_e^{\rm surv}) = \gaussian(S2| g_2 n_{e^{\rm surv}}, \sigma_{S2})  $, with a representative
width $\sigma_{S2} = 7 \sqrt{n_e^{\rm surv}}$~\cite{Aprile:2013blg} and a gain factor
$g_2 = 33, 20$, and $27$~PE/$e^-$ for XENON1T, XENON100, and XENON10, respectively. For silicon, we use $dR/dn_e=\int dE_e P(E_e|n_e)dR/dE_e $ where  $P(E_e|n_e)=\delta(1+$Floor$[E_e/3.8~{\rm eV}]-n_e)$. 
We reinterpret the data and efficiencies for XENON10~\cite{Angle:2011th}, XENON100~\cite{Aprile:2016wwo}, XENON1T~\cite{Aprile:2019xxb}, and SENSEI~\cite{Abramoff:2019dfb} to obtain the exclusion boundaries in Fig.~\ref{fig:constraints}. We leave to future work calculating the limits from DarkSide-50~\cite{Agnes:2018oej}, CDMS-HVeV~\cite{Agnese:2018col}, and DAMIC at SNOLAB~\cite{Aguilar-Arevalo:2019wdi}. 
We also show the 90\% C.L.~projections for a future 100 g-year silicon detector (like SENSEI) with a single-electron threshold, a dark-count of $10^6$ events for  $n_e=1$ and no background events for $n_e\ge2$; and a 100 kg-year xenon detector (like LBECA) with a two-electron threshold and assuming only neutrino induced backgrounds~\cite{Tiffenberg:2017aac,Essig:2018tss,Battaglieri:2017aum}.  
 The new constraint 
 from XENON10 extends sensitivity to DM-nuclear scattering down to $m_\chi \sim 5\, \MeV$, and future experiments will reach $\sim 500\,\keV$. 
 We also compare the DM-electron scattering limits with the Migdal effect, assuming a dark photon mediator, and converting the limits and projections from $\overline\sigma_e$ to $\overline\sigma_n$ using Eq.~(\ref{eq:darkphoton}). For contact interactions, the DM-electron scattering limits are stronger for smaller DM masses, but weaker for heavier DM masses, than the Migdal limits. For light mediators, the DM-electron scattering dominates for all masses. 

\paragraph{\textbf{Conclusions.}} 

The prompt electron ionization 
signal that may accompany DM-nuclear scattering, ``Migdal effect'', allows to extend the sensitivity of noble liquid and semiconductor DM detectors into the MeV mass region. In this work, we show that the theoretical description of  the process is closely related to DM-electron scattering, and set the lowest DM-mass constraints on DM-nuclear scattering using XENON10 and XENON100 data. 
Finally, we take the first step towards a concise formulation of the Migdal effect in semiconductors and demonstrate the future potential for both, upcoming noble liquid and semiconductor experiments.

\paragraph{\textbf{Acknowledgements.}} We thank R.~Budnik and R.~Lang for correspondence on the XENON1T results. R.E.~and M.S.'s~work in this paper is supported by
DoE Grant DE-SC0017938. R.E.~also acknowledges support
from the US-Israel Binational Science Foundation
under Grant No.~2016153, from the Heising-Simons Foundation
under Grant No.~79921, from a subaward for the
DOE Grant No.~DE-SC0018952, and from Simons Investigator
Award 623940. J.P.~is supported by the ‘New Frontiers’ program of the Austrian Academy of Sciences. We thank the Erwin Schr\"odinger International Institute for hospitality while this work was completed. T.-T.Y.~also thanks the hospitality of the CERN theory department where part of this work was completed.

\paragraph*{\textbf{Note Added}}

During the completion of this work, Ref.~\cite{Baxter:2019pnz} appeared which investigates similar topics for the isolated atom case. 

\bibliography{main.bbl}

\begin{thebibliography}{68}%
\makeatletter
\providecommand \@ifxundefined [1]{%
 \@ifx{#1\undefined}
}%
\providecommand \@ifnum [1]{%
 \ifnum #1\expandafter \@firstoftwo
 \else \expandafter \@secondoftwo
 \fi
}%
\providecommand \@ifx [1]{%
 \ifx #1\expandafter \@firstoftwo
 \else \expandafter \@secondoftwo
 \fi
}%
\providecommand \natexlab [1]{#1}%
\providecommand \enquote  [1]{``#1''}%
\providecommand \bibnamefont  [1]{#1}%
\providecommand \bibfnamefont [1]{#1}%
\providecommand \citenamefont [1]{#1}%
\providecommand \href@noop [0]{\@secondoftwo}%
\providecommand \href [0]{\begingroup \@sanitize@url \@href}%
\providecommand \@href[1]{\@@startlink{#1}\@@href}%
\providecommand \@@href[1]{\endgroup#1\@@endlink}%
\providecommand \@sanitize@url [0]{\catcode `\\12\catcode `\$12\catcode
  `\&12\catcode `\#12\catcode `\^12\catcode `\_12\catcode `\%12\relax}%
\providecommand \@@startlink[1]{}%
\providecommand \@@endlink[0]{}%
\providecommand \url  [0]{\begingroup\@sanitize@url \@url }%
\providecommand \@url [1]{\endgroup\@href {#1}{\urlprefix }}%
\providecommand \urlprefix  [0]{URL }%
\providecommand \Eprint [0]{\href }%
\providecommand \doibase [0]{http://dx.doi.org/}%
\providecommand \selectlanguage [0]{\@gobble}%
\providecommand \bibinfo  [0]{\@secondoftwo}%
\providecommand \bibfield  [0]{\@secondoftwo}%
\providecommand \translation [1]{[#1]}%
\providecommand \BibitemOpen [0]{}%
\providecommand \bibitemStop [0]{}%
\providecommand \bibitemNoStop [0]{.\EOS\space}%
\providecommand \EOS [0]{\spacefactor3000\relax}%
\providecommand \BibitemShut  [1]{\csname bibitem#1\endcsname}%
\let\auto@bib@innerbib\@empty
\bibitem [{\citenamefont {Kouvaris}\ and\ \citenamefont
  {Pradler}(2017)}]{Kouvaris:2016afs}%
  \BibitemOpen
  \bibfield  {author} {\bibinfo {author} {\bibfnamefont {C.}~\bibnamefont
  {Kouvaris}}\ and\ \bibinfo {author} {\bibfnamefont {J.}~\bibnamefont
  {Pradler}},\ }\href {\doibase 10.1103/PhysRevLett.118.031803} {\bibfield
  {journal} {\bibinfo  {journal} {Phys. Rev. Lett.}\ }\textbf {\bibinfo
  {volume} {118}},\ \bibinfo {pages} {031803} (\bibinfo {year} {2017})},\
  \Eprint {http://arxiv.org/abs/1607.01789} {arXiv:1607.01789 [hep-ph]}
  \BibitemShut {NoStop}%
\bibitem [{\citenamefont {Ibe}\ \emph {et~al.}(2018)\citenamefont {Ibe},
  \citenamefont {Nakano}, \citenamefont {Shoji},\ and\ \citenamefont
  {Suzuki}}]{Ibe:2017yqa}%
  \BibitemOpen
  \bibfield  {author} {\bibinfo {author} {\bibfnamefont {M.}~\bibnamefont
  {Ibe}}, \bibinfo {author} {\bibfnamefont {W.}~\bibnamefont {Nakano}},
  \bibinfo {author} {\bibfnamefont {Y.}~\bibnamefont {Shoji}}, \ and\ \bibinfo
  {author} {\bibfnamefont {K.}~\bibnamefont {Suzuki}},\ }\href {\doibase
  10.1007/JHEP03(2018)194} {\bibfield  {journal} {\bibinfo  {journal} {JHEP}\
  }\textbf {\bibinfo {volume} {03}},\ \bibinfo {pages} {194} (\bibinfo {year}
  {2018})},\ \Eprint {http://arxiv.org/abs/1707.07258} {arXiv:1707.07258
  [hep-ph]} \BibitemShut {NoStop}%
\bibitem [{\citenamefont {Essig}\ \emph
  {et~al.}(2012{\natexlab{a}})\citenamefont {Essig}, \citenamefont {Mardon},\
  and\ \citenamefont {Volansky}}]{Essig:2011nj}%
  \BibitemOpen
  \bibfield  {author} {\bibinfo {author} {\bibfnamefont {R.}~\bibnamefont
  {Essig}}, \bibinfo {author} {\bibfnamefont {J.}~\bibnamefont {Mardon}}, \
  and\ \bibinfo {author} {\bibfnamefont {T.}~\bibnamefont {Volansky}},\ }\href
  {\doibase 10.1103/PhysRevD.85.076007} {\bibfield  {journal} {\bibinfo
  {journal} {Phys. Rev.}\ }\textbf {\bibinfo {volume} {D85}},\ \bibinfo {pages}
  {076007} (\bibinfo {year} {2012}{\natexlab{a}})},\ \Eprint
  {http://arxiv.org/abs/1108.5383} {arXiv:1108.5383 [hep-ph]} \BibitemShut
  {NoStop}%
\bibitem [{\citenamefont {Akerib}\ \emph {et~al.}(2019)\citenamefont {Akerib}
  \emph {et~al.}}]{Akerib:2018hck}%
  \BibitemOpen
  \bibfield  {author} {\bibinfo {author} {\bibfnamefont {D.~S.}\ \bibnamefont
  {Akerib}} \emph {et~al.} (\bibinfo {collaboration} {LUX}),\ }\href {\doibase
  10.1103/PhysRevLett.122.131301} {\bibfield  {journal} {\bibinfo  {journal}
  {Phys. Rev. Lett.}\ }\textbf {\bibinfo {volume} {122}},\ \bibinfo {pages}
  {131301} (\bibinfo {year} {2019})},\ \Eprint
  {http://arxiv.org/abs/1811.11241} {arXiv:1811.11241 [astro-ph.CO]}
  \BibitemShut {NoStop}%
\bibitem [{\citenamefont {Armengaud}\ \emph {et~al.}(2019)\citenamefont
  {Armengaud} \emph {et~al.}}]{Armengaud:2019kfj}%
  \BibitemOpen
  \bibfield  {author} {\bibinfo {author} {\bibfnamefont {E.}~\bibnamefont
  {Armengaud}} \emph {et~al.} (\bibinfo {collaboration} {EDELWEISS}),\ }\href
  {\doibase 10.1103/PhysRevD.99.082003} {\bibfield  {journal} {\bibinfo
  {journal} {Phys. Rev.}\ }\textbf {\bibinfo {volume} {D99}},\ \bibinfo {pages}
  {082003} (\bibinfo {year} {2019})},\ \Eprint
  {http://arxiv.org/abs/1901.03588} {arXiv:1901.03588 [astro-ph.GA]}
  \BibitemShut {NoStop}%
\bibitem [{\citenamefont {Liu}\ \emph {et~al.}(2019)\citenamefont {Liu} \emph
  {et~al.}}]{Liu:2019kzq}%
  \BibitemOpen
  \bibfield  {author} {\bibinfo {author} {\bibfnamefont {Z.~Z.}\ \bibnamefont
  {Liu}} \emph {et~al.} (\bibinfo {collaboration} {CDEX}),\ }\href@noop {} {\
  (\bibinfo {year} {2019})},\ \Eprint {http://arxiv.org/abs/1905.00354}
  {arXiv:1905.00354 [hep-ex]} \BibitemShut {NoStop}%
\bibitem [{\citenamefont {Aprile}\ \emph
  {et~al.}(2019{\natexlab{a}})\citenamefont {Aprile} \emph
  {et~al.}}]{Aprile:2019jmx}%
  \BibitemOpen
  \bibfield  {author} {\bibinfo {author} {\bibfnamefont {E.}~\bibnamefont
  {Aprile}} \emph {et~al.} (\bibinfo {collaboration} {XENON}),\ }\href@noop {}
  {\  (\bibinfo {year} {2019}{\natexlab{a}})},\ \Eprint
  {http://arxiv.org/abs/1907.12771} {arXiv:1907.12771 [hep-ex]} \BibitemShut
  {NoStop}%
\bibitem [{\citenamefont {Essig}\ \emph
  {et~al.}(2012{\natexlab{b}})\citenamefont {Essig}, \citenamefont
  {Manalaysay}, \citenamefont {Mardon}, \citenamefont {Sorensen},\ and\
  \citenamefont {Volansky}}]{Essig:2012yx}%
  \BibitemOpen
  \bibfield  {author} {\bibinfo {author} {\bibfnamefont {R.}~\bibnamefont
  {Essig}}, \bibinfo {author} {\bibfnamefont {A.}~\bibnamefont {Manalaysay}},
  \bibinfo {author} {\bibfnamefont {J.}~\bibnamefont {Mardon}}, \bibinfo
  {author} {\bibfnamefont {P.}~\bibnamefont {Sorensen}}, \ and\ \bibinfo
  {author} {\bibfnamefont {T.}~\bibnamefont {Volansky}},\ }\href {\doibase
  10.1103/PhysRevLett.109.021301} {\bibfield  {journal} {\bibinfo  {journal}
  {Phys. Rev. Lett.}\ }\textbf {\bibinfo {volume} {109}},\ \bibinfo {pages}
  {021301} (\bibinfo {year} {2012}{\natexlab{b}})},\ \Eprint
  {http://arxiv.org/abs/1206.2644} {arXiv:1206.2644 [astro-ph.CO]} \BibitemShut
  {NoStop}%
\bibitem [{\citenamefont {Essig}\ \emph
  {et~al.}(2017{\natexlab{a}})\citenamefont {Essig}, \citenamefont {Volansky},\
  and\ \citenamefont {Yu}}]{Essig:2017kqs}%
  \BibitemOpen
  \bibfield  {author} {\bibinfo {author} {\bibfnamefont {R.}~\bibnamefont
  {Essig}}, \bibinfo {author} {\bibfnamefont {T.}~\bibnamefont {Volansky}}, \
  and\ \bibinfo {author} {\bibfnamefont {T.-T.}\ \bibnamefont {Yu}},\ }\href
  {\doibase 10.1103/PhysRevD.96.043017} {\bibfield  {journal} {\bibinfo
  {journal} {Phys. Rev.}\ }\textbf {\bibinfo {volume} {D96}},\ \bibinfo {pages}
  {043017} (\bibinfo {year} {2017}{\natexlab{a}})},\ \Eprint
  {http://arxiv.org/abs/1703.00910} {arXiv:1703.00910 [hep-ph]} \BibitemShut
  {NoStop}%
\bibitem [{\citenamefont {Tiffenberg}\ \emph {et~al.}(2017)\citenamefont
  {Tiffenberg}, \citenamefont {Sofo-Haro}, \citenamefont {Drlica-Wagner},
  \citenamefont {Essig}, \citenamefont {Guardincerri}, \citenamefont {Holland},
  \citenamefont {Volansky},\ and\ \citenamefont {Yu}}]{Tiffenberg:2017aac}%
  \BibitemOpen
  \bibfield  {author} {\bibinfo {author} {\bibfnamefont {J.}~\bibnamefont
  {Tiffenberg}}, \bibinfo {author} {\bibfnamefont {M.}~\bibnamefont
  {Sofo-Haro}}, \bibinfo {author} {\bibfnamefont {A.}~\bibnamefont
  {Drlica-Wagner}}, \bibinfo {author} {\bibfnamefont {R.}~\bibnamefont
  {Essig}}, \bibinfo {author} {\bibfnamefont {Y.}~\bibnamefont {Guardincerri}},
  \bibinfo {author} {\bibfnamefont {S.}~\bibnamefont {Holland}}, \bibinfo
  {author} {\bibfnamefont {T.}~\bibnamefont {Volansky}}, \ and\ \bibinfo
  {author} {\bibfnamefont {T.-T.}\ \bibnamefont {Yu}} (\bibinfo {collaboration}
  {SENSEI}),\ }\href {\doibase 10.1103/PhysRevLett.119.131802} {\bibfield
  {journal} {\bibinfo  {journal} {Phys. Rev. Lett.}\ }\textbf {\bibinfo
  {volume} {119}},\ \bibinfo {pages} {131802} (\bibinfo {year} {2017})},\
  \Eprint {http://arxiv.org/abs/1706.00028} {arXiv:1706.00028
  [physics.ins-det]} \BibitemShut {NoStop}%
\bibitem [{\citenamefont {Romani}\ \emph {et~al.}(2018)\citenamefont {Romani}
  \emph {et~al.}}]{Romani:2017iwi}%
  \BibitemOpen
  \bibfield  {author} {\bibinfo {author} {\bibfnamefont {R.~K.}\ \bibnamefont
  {Romani}} \emph {et~al.},\ }\href {\doibase 10.1063/1.5010699} {\bibfield
  {journal} {\bibinfo  {journal} {Appl. Phys. Lett.}\ }\textbf {\bibinfo
  {volume} {112}},\ \bibinfo {pages} {043501} (\bibinfo {year} {2018})},\
  \Eprint {http://arxiv.org/abs/1710.09335} {arXiv:1710.09335
  [physics.ins-det]} \BibitemShut {NoStop}%
\bibitem [{\citenamefont {Crisler}\ \emph {et~al.}(2018)\citenamefont
  {Crisler}, \citenamefont {Essig}, \citenamefont {Estrada}, \citenamefont
  {Fernandez}, \citenamefont {Tiffenberg}, \citenamefont {Sofo~haro},
  \citenamefont {Volansky},\ and\ \citenamefont {Yu}}]{Crisler:2018gci}%
  \BibitemOpen
  \bibfield  {author} {\bibinfo {author} {\bibfnamefont {M.}~\bibnamefont
  {Crisler}}, \bibinfo {author} {\bibfnamefont {R.}~\bibnamefont {Essig}},
  \bibinfo {author} {\bibfnamefont {J.}~\bibnamefont {Estrada}}, \bibinfo
  {author} {\bibfnamefont {G.}~\bibnamefont {Fernandez}}, \bibinfo {author}
  {\bibfnamefont {J.}~\bibnamefont {Tiffenberg}}, \bibinfo {author}
  {\bibfnamefont {M.}~\bibnamefont {Sofo~haro}}, \bibinfo {author}
  {\bibfnamefont {T.}~\bibnamefont {Volansky}}, \ and\ \bibinfo {author}
  {\bibfnamefont {T.-T.}\ \bibnamefont {Yu}} (\bibinfo {collaboration}
  {SENSEI}),\ }\href@noop {} {\  (\bibinfo {year} {2018})},\ \Eprint
  {http://arxiv.org/abs/1804.00088} {arXiv:1804.00088 [hep-ex]} \BibitemShut
  {NoStop}%
\bibitem [{\citenamefont {Agnese}\ \emph {et~al.}(2018)\citenamefont {Agnese}
  \emph {et~al.}}]{Agnese:2018col}%
  \BibitemOpen
  \bibfield  {author} {\bibinfo {author} {\bibfnamefont {R.}~\bibnamefont
  {Agnese}} \emph {et~al.} (\bibinfo {collaboration} {SuperCDMS}),\ }\href
  {\doibase 10.1103/PhysRevLett.122.069901, 10.1103/PhysRevLett.121.051301}
  {\bibfield  {journal} {\bibinfo  {journal} {Phys. Rev. Lett.}\ }\textbf
  {\bibinfo {volume} {121}},\ \bibinfo {pages} {051301} (\bibinfo {year}
  {2018})},\ \bibinfo {note} {[erratum: Phys. Rev.
  Lett.122,no.6,069901(2019)]},\ \Eprint {http://arxiv.org/abs/1804.10697}
  {arXiv:1804.10697 [hep-ex]} \BibitemShut {NoStop}%
\bibitem [{\citenamefont {Agnes}\ \emph
  {et~al.}(2018{\natexlab{a}})\citenamefont {Agnes} \emph
  {et~al.}}]{Agnes:2018oej}%
  \BibitemOpen
  \bibfield  {author} {\bibinfo {author} {\bibfnamefont {P.}~\bibnamefont
  {Agnes}} \emph {et~al.} (\bibinfo {collaboration} {DarkSide}),\ }\href
  {\doibase 10.1103/PhysRevLett.121.111303} {\bibfield  {journal} {\bibinfo
  {journal} {Phys. Rev. Lett.}\ }\textbf {\bibinfo {volume} {121}},\ \bibinfo
  {pages} {111303} (\bibinfo {year} {2018}{\natexlab{a}})},\ \Eprint
  {http://arxiv.org/abs/1802.06998} {arXiv:1802.06998 [astro-ph.CO]}
  \BibitemShut {NoStop}%
\bibitem [{\citenamefont {Abramoff}\ \emph {et~al.}(2019)\citenamefont
  {Abramoff} \emph {et~al.}}]{Abramoff:2019dfb}%
  \BibitemOpen
  \bibfield  {author} {\bibinfo {author} {\bibfnamefont {O.}~\bibnamefont
  {Abramoff}} \emph {et~al.} (\bibinfo {collaboration} {SENSEI}),\ }\href
  {\doibase 10.1103/PhysRevLett.122.161801} {\bibfield  {journal} {\bibinfo
  {journal} {Phys. Rev. Lett.}\ }\textbf {\bibinfo {volume} {122}},\ \bibinfo
  {pages} {161801} (\bibinfo {year} {2019})},\ \Eprint
  {http://arxiv.org/abs/1901.10478} {arXiv:1901.10478 [hep-ex]} \BibitemShut
  {NoStop}%
\bibitem [{\citenamefont {Aguilar-Arevalo}\ \emph {et~al.}(2019)\citenamefont
  {Aguilar-Arevalo} \emph {et~al.}}]{Aguilar-Arevalo:2019wdi}%
  \BibitemOpen
  \bibfield  {author} {\bibinfo {author} {\bibfnamefont {A.}~\bibnamefont
  {Aguilar-Arevalo}} \emph {et~al.} (\bibinfo {collaboration} {DAMIC}),\
  }\href@noop {} {\  (\bibinfo {year} {2019})},\ \Eprint
  {http://arxiv.org/abs/1907.12628} {arXiv:1907.12628 [astro-ph.CO]}
  \BibitemShut {NoStop}%
\bibitem [{\citenamefont {Settimo}(2018)}]{Settimo:2018qcm}%
  \BibitemOpen
  \bibfield  {author} {\bibinfo {author} {\bibfnamefont {M.}~\bibnamefont
  {Settimo}} (\bibinfo {collaboration} {DAMIC}),\ }in\ \href@noop {} {\emph
  {\bibinfo {booktitle} {{53rd Rencontres de Moriond on QCD and High Energy
  Interactions (Moriond QCD 2018) La Thuile, Italy, March 17-24, 2018}}}}\
  (\bibinfo {year} {2018})\ \Eprint {http://arxiv.org/abs/1805.10001}
  {arXiv:1805.10001 [astro-ph.IM]} \BibitemShut {NoStop}%
\bibitem [{\citenamefont {Vergados}\ and\ \citenamefont
  {Ejiri}(2005)}]{Vergados:2004bm}%
  \BibitemOpen
  \bibfield  {author} {\bibinfo {author} {\bibfnamefont {J.~D.}\ \bibnamefont
  {Vergados}}\ and\ \bibinfo {author} {\bibfnamefont {H.}~\bibnamefont
  {Ejiri}},\ }\href {\doibase 10.1016/j.physletb.2004.11.085} {\bibfield
  {journal} {\bibinfo  {journal} {Phys. Lett.}\ }\textbf {\bibinfo {volume}
  {B606}},\ \bibinfo {pages} {313} (\bibinfo {year} {2005})},\ \Eprint
  {http://arxiv.org/abs/hep-ph/0401151} {arXiv:hep-ph/0401151 [hep-ph]}
  \BibitemShut {NoStop}%
\bibitem [{\citenamefont {Moustakidis}\ \emph {et~al.}(2005)\citenamefont
  {Moustakidis}, \citenamefont {Vergados},\ and\ \citenamefont
  {Ejiri}}]{Moustakidis:2005gx}%
  \BibitemOpen
  \bibfield  {author} {\bibinfo {author} {\bibfnamefont {C.~C.}\ \bibnamefont
  {Moustakidis}}, \bibinfo {author} {\bibfnamefont {J.~D.}\ \bibnamefont
  {Vergados}}, \ and\ \bibinfo {author} {\bibfnamefont {H.}~\bibnamefont
  {Ejiri}},\ }\href {\doibase 10.1016/j.nuclphysb.2005.08.033} {\bibfield
  {journal} {\bibinfo  {journal} {Nucl. Phys.}\ }\textbf {\bibinfo {volume}
  {B727}},\ \bibinfo {pages} {406} (\bibinfo {year} {2005})},\ \Eprint
  {http://arxiv.org/abs/hep-ph/0507123} {arXiv:hep-ph/0507123 [hep-ph]}
  \BibitemShut {NoStop}%
\bibitem [{\citenamefont {Ejiri}\ \emph {et~al.}(2006)\citenamefont {Ejiri},
  \citenamefont {Moustakidis},\ and\ \citenamefont {Vergados}}]{Ejiri:2005aj}%
  \BibitemOpen
  \bibfield  {author} {\bibinfo {author} {\bibfnamefont {H.}~\bibnamefont
  {Ejiri}}, \bibinfo {author} {\bibfnamefont {C.~C.}\ \bibnamefont
  {Moustakidis}}, \ and\ \bibinfo {author} {\bibfnamefont {J.~D.}\ \bibnamefont
  {Vergados}},\ }\href {\doibase 10.1016/j.physletb.2006.03.037} {\bibfield
  {journal} {\bibinfo  {journal} {Phys. Lett.}\ }\textbf {\bibinfo {volume}
  {B639}},\ \bibinfo {pages} {218} (\bibinfo {year} {2006})},\ \Eprint
  {http://arxiv.org/abs/hep-ph/0510042} {arXiv:hep-ph/0510042 [hep-ph]}
  \BibitemShut {NoStop}%
\bibitem [{\citenamefont {Bernabei}\ \emph {et~al.}(2007)\citenamefont
  {Bernabei} \emph {et~al.}}]{Bernabei:2007jz}%
  \BibitemOpen
  \bibfield  {author} {\bibinfo {author} {\bibfnamefont {R.}~\bibnamefont
  {Bernabei}} \emph {et~al.},\ }\href {\doibase 10.1142/S0217751X07037093}
  {\bibfield  {journal} {\bibinfo  {journal} {Int. J. Mod. Phys.}\ }\textbf
  {\bibinfo {volume} {A22}},\ \bibinfo {pages} {3155} (\bibinfo {year}
  {2007})},\ \Eprint {http://arxiv.org/abs/0706.1421} {arXiv:0706.1421
  [astro-ph]} \BibitemShut {NoStop}%
\bibitem [{\citenamefont {Essig}\ \emph {et~al.}(2016)\citenamefont {Essig},
  \citenamefont {Fernandez-Serra}, \citenamefont {Mardon}, \citenamefont
  {Soto}, \citenamefont {Volansky},\ and\ \citenamefont {Yu}}]{Essig:2015cda}%
  \BibitemOpen
  \bibfield  {author} {\bibinfo {author} {\bibfnamefont {R.}~\bibnamefont
  {Essig}}, \bibinfo {author} {\bibfnamefont {M.}~\bibnamefont
  {Fernandez-Serra}}, \bibinfo {author} {\bibfnamefont {J.}~\bibnamefont
  {Mardon}}, \bibinfo {author} {\bibfnamefont {A.}~\bibnamefont {Soto}},
  \bibinfo {author} {\bibfnamefont {T.}~\bibnamefont {Volansky}}, \ and\
  \bibinfo {author} {\bibfnamefont {T.-T.}\ \bibnamefont {Yu}},\ }\href
  {\doibase 10.1007/JHEP05(2016)046} {\bibfield  {journal} {\bibinfo  {journal}
  {JHEP}\ }\textbf {\bibinfo {volume} {05}},\ \bibinfo {pages} {046} (\bibinfo
  {year} {2016})},\ \Eprint {http://arxiv.org/abs/1509.01598} {arXiv:1509.01598
  [hep-ph]} \BibitemShut {NoStop}%
\bibitem [{\citenamefont {Graham}\ \emph {et~al.}(2012)\citenamefont {Graham},
  \citenamefont {Kaplan}, \citenamefont {Rajendran},\ and\ \citenamefont
  {Walters}}]{Graham:2012su}%
  \BibitemOpen
  \bibfield  {author} {\bibinfo {author} {\bibfnamefont {P.~W.}\ \bibnamefont
  {Graham}}, \bibinfo {author} {\bibfnamefont {D.~E.}\ \bibnamefont {Kaplan}},
  \bibinfo {author} {\bibfnamefont {S.}~\bibnamefont {Rajendran}}, \ and\
  \bibinfo {author} {\bibfnamefont {M.~T.}\ \bibnamefont {Walters}},\ }\href
  {\doibase 10.1016/j.dark.2012.09.001} {\bibfield  {journal} {\bibinfo
  {journal} {Phys. Dark Univ.}\ }\textbf {\bibinfo {volume} {1}},\ \bibinfo
  {pages} {32} (\bibinfo {year} {2012})},\ \Eprint
  {http://arxiv.org/abs/1203.2531} {arXiv:1203.2531 [hep-ph]} \BibitemShut
  {NoStop}%
\bibitem [{\citenamefont {An}\ \emph {et~al.}(2015)\citenamefont {An},
  \citenamefont {Pospelov}, \citenamefont {Pradler},\ and\ \citenamefont
  {Ritz}}]{An:2014twa}%
  \BibitemOpen
  \bibfield  {author} {\bibinfo {author} {\bibfnamefont {H.}~\bibnamefont
  {An}}, \bibinfo {author} {\bibfnamefont {M.}~\bibnamefont {Pospelov}},
  \bibinfo {author} {\bibfnamefont {J.}~\bibnamefont {Pradler}}, \ and\
  \bibinfo {author} {\bibfnamefont {A.}~\bibnamefont {Ritz}},\ }\href {\doibase
  10.1016/j.physletb.2015.06.018} {\bibfield  {journal} {\bibinfo  {journal}
  {Phys. Lett.}\ }\textbf {\bibinfo {volume} {B747}},\ \bibinfo {pages} {331}
  (\bibinfo {year} {2015})},\ \Eprint {http://arxiv.org/abs/1412.8378}
  {arXiv:1412.8378 [hep-ph]} \BibitemShut {NoStop}%
\bibitem [{\citenamefont {Aprile}\ \emph
  {et~al.}(2014{\natexlab{a}})\citenamefont {Aprile} \emph
  {et~al.}}]{Aprile:2014eoa}%
  \BibitemOpen
  \bibfield  {author} {\bibinfo {author} {\bibfnamefont {E.}~\bibnamefont
  {Aprile}} \emph {et~al.} (\bibinfo {collaboration} {XENON100}),\ }\href
  {\doibase 10.1103/PhysRevD.90.062009, 10.1103/PhysRevD.95.029904} {\bibfield
  {journal} {\bibinfo  {journal} {Phys. Rev.}\ }\textbf {\bibinfo {volume}
  {D90}},\ \bibinfo {pages} {062009} (\bibinfo {year} {2014}{\natexlab{a}})},\
  \bibinfo {note} {[Erratum: Phys. Rev.D95,no.2,029904(2017)]},\ \Eprint
  {http://arxiv.org/abs/1404.1455} {arXiv:1404.1455 [astro-ph.CO]} \BibitemShut
  {NoStop}%
\bibitem [{\citenamefont {Lee}\ \emph {et~al.}(2015)\citenamefont {Lee},
  \citenamefont {Lisanti}, \citenamefont {Mishra-Sharma},\ and\ \citenamefont
  {Safdi}}]{Lee:2015qva}%
  \BibitemOpen
  \bibfield  {author} {\bibinfo {author} {\bibfnamefont {S.~K.}\ \bibnamefont
  {Lee}}, \bibinfo {author} {\bibfnamefont {M.}~\bibnamefont {Lisanti}},
  \bibinfo {author} {\bibfnamefont {S.}~\bibnamefont {Mishra-Sharma}}, \ and\
  \bibinfo {author} {\bibfnamefont {B.~R.}\ \bibnamefont {Safdi}},\ }\href
  {\doibase 10.1103/PhysRevD.92.083517} {\bibfield  {journal} {\bibinfo
  {journal} {Phys. Rev.}\ }\textbf {\bibinfo {volume} {D92}},\ \bibinfo {pages}
  {083517} (\bibinfo {year} {2015})},\ \Eprint
  {http://arxiv.org/abs/1508.07361} {arXiv:1508.07361 [hep-ph]} \BibitemShut
  {NoStop}%
\bibitem [{\citenamefont {Hochberg}\ \emph
  {et~al.}(2016{\natexlab{a}})\citenamefont {Hochberg}, \citenamefont {Zhao},\
  and\ \citenamefont {Zurek}}]{Hochberg:2015pha}%
  \BibitemOpen
  \bibfield  {author} {\bibinfo {author} {\bibfnamefont {Y.}~\bibnamefont
  {Hochberg}}, \bibinfo {author} {\bibfnamefont {Y.}~\bibnamefont {Zhao}}, \
  and\ \bibinfo {author} {\bibfnamefont {K.~M.}\ \bibnamefont {Zurek}},\ }\href
  {\doibase 10.1103/PhysRevLett.116.011301} {\bibfield  {journal} {\bibinfo
  {journal} {Phys. Rev. Lett.}\ }\textbf {\bibinfo {volume} {116}},\ \bibinfo
  {pages} {011301} (\bibinfo {year} {2016}{\natexlab{a}})},\ \Eprint
  {http://arxiv.org/abs/1504.07237} {arXiv:1504.07237 [hep-ph]} \BibitemShut
  {NoStop}%
\bibitem [{\citenamefont {Hochberg}\ \emph
  {et~al.}(2016{\natexlab{b}})\citenamefont {Hochberg}, \citenamefont {Pyle},
  \citenamefont {Zhao},\ and\ \citenamefont {Zurek}}]{Hochberg:2015fth}%
  \BibitemOpen
  \bibfield  {author} {\bibinfo {author} {\bibfnamefont {Y.}~\bibnamefont
  {Hochberg}}, \bibinfo {author} {\bibfnamefont {M.}~\bibnamefont {Pyle}},
  \bibinfo {author} {\bibfnamefont {Y.}~\bibnamefont {Zhao}}, \ and\ \bibinfo
  {author} {\bibfnamefont {K.~M.}\ \bibnamefont {Zurek}},\ }\href {\doibase
  10.1007/JHEP08(2016)057} {\bibfield  {journal} {\bibinfo  {journal} {JHEP}\
  }\textbf {\bibinfo {volume} {08}},\ \bibinfo {pages} {057} (\bibinfo {year}
  {2016}{\natexlab{b}})},\ \Eprint {http://arxiv.org/abs/1512.04533}
  {arXiv:1512.04533 [hep-ph]} \BibitemShut {NoStop}%
\bibitem [{\citenamefont {Aguilar-Arevalo}\ \emph {et~al.}(2017)\citenamefont
  {Aguilar-Arevalo} \emph {et~al.}}]{Aguilar-Arevalo:2016zop}%
  \BibitemOpen
  \bibfield  {author} {\bibinfo {author} {\bibfnamefont {A.}~\bibnamefont
  {Aguilar-Arevalo}} \emph {et~al.} (\bibinfo {collaboration} {DAMIC}),\ }\href
  {\doibase 10.1103/PhysRevLett.118.141803} {\bibfield  {journal} {\bibinfo
  {journal} {Phys. Rev. Lett.}\ }\textbf {\bibinfo {volume} {118}},\ \bibinfo
  {pages} {141803} (\bibinfo {year} {2017})},\ \Eprint
  {http://arxiv.org/abs/1611.03066} {arXiv:1611.03066 [astro-ph.CO]}
  \BibitemShut {NoStop}%
\bibitem [{\citenamefont {Bloch}\ \emph {et~al.}(2017)\citenamefont {Bloch},
  \citenamefont {Essig}, \citenamefont {Tobioka}, \citenamefont {Volansky},\
  and\ \citenamefont {Yu}}]{Bloch:2016sjj}%
  \BibitemOpen
  \bibfield  {author} {\bibinfo {author} {\bibfnamefont {I.~M.}\ \bibnamefont
  {Bloch}}, \bibinfo {author} {\bibfnamefont {R.}~\bibnamefont {Essig}},
  \bibinfo {author} {\bibfnamefont {K.}~\bibnamefont {Tobioka}}, \bibinfo
  {author} {\bibfnamefont {T.}~\bibnamefont {Volansky}}, \ and\ \bibinfo
  {author} {\bibfnamefont {T.-T.}\ \bibnamefont {Yu}},\ }\href {\doibase
  10.1007/JHEP06(2017)087} {\bibfield  {journal} {\bibinfo  {journal} {JHEP}\
  }\textbf {\bibinfo {volume} {06}},\ \bibinfo {pages} {087} (\bibinfo {year}
  {2017})},\ \Eprint {http://arxiv.org/abs/1608.02123} {arXiv:1608.02123
  [hep-ph]} \BibitemShut {NoStop}%
\bibitem [{\citenamefont {Cavoto}\ \emph {et~al.}(2016)\citenamefont {Cavoto},
  \citenamefont {Cirillo}, \citenamefont {Cocina}, \citenamefont {Ferretti},\
  and\ \citenamefont {Polosa}}]{Cavoto:2016lqo}%
  \BibitemOpen
  \bibfield  {author} {\bibinfo {author} {\bibfnamefont {G.}~\bibnamefont
  {Cavoto}}, \bibinfo {author} {\bibfnamefont {E.~N.~M.}\ \bibnamefont
  {Cirillo}}, \bibinfo {author} {\bibfnamefont {F.}~\bibnamefont {Cocina}},
  \bibinfo {author} {\bibfnamefont {J.}~\bibnamefont {Ferretti}}, \ and\
  \bibinfo {author} {\bibfnamefont {A.~D.}\ \bibnamefont {Polosa}},\ }\href
  {\doibase 10.1140/epjc/s10052-016-4193-7} {\bibfield  {journal} {\bibinfo
  {journal} {Eur. Phys. J.}\ }\textbf {\bibinfo {volume} {C76}},\ \bibinfo
  {pages} {349} (\bibinfo {year} {2016})},\ \Eprint
  {http://arxiv.org/abs/1602.03216} {arXiv:1602.03216 [physics.ins-det]}
  \BibitemShut {NoStop}%
\bibitem [{\citenamefont {Derenzo}\ \emph {et~al.}(2017)\citenamefont
  {Derenzo}, \citenamefont {Essig}, \citenamefont {Massari}, \citenamefont
  {Soto},\ and\ \citenamefont {Yu}}]{Derenzo:2016fse}%
  \BibitemOpen
  \bibfield  {author} {\bibinfo {author} {\bibfnamefont {S.}~\bibnamefont
  {Derenzo}}, \bibinfo {author} {\bibfnamefont {R.}~\bibnamefont {Essig}},
  \bibinfo {author} {\bibfnamefont {A.}~\bibnamefont {Massari}}, \bibinfo
  {author} {\bibfnamefont {A.}~\bibnamefont {Soto}}, \ and\ \bibinfo {author}
  {\bibfnamefont {T.-T.}\ \bibnamefont {Yu}},\ }\href {\doibase
  10.1103/PhysRevD.96.016026} {\bibfield  {journal} {\bibinfo  {journal} {Phys.
  Rev.}\ }\textbf {\bibinfo {volume} {D96}},\ \bibinfo {pages} {016026}
  (\bibinfo {year} {2017})},\ \Eprint {http://arxiv.org/abs/1607.01009}
  {arXiv:1607.01009 [hep-ph]} \BibitemShut {NoStop}%
\bibitem [{\citenamefont {Essig}\ \emph
  {et~al.}(2017{\natexlab{b}})\citenamefont {Essig}, \citenamefont {Mardon},
  \citenamefont {Slone},\ and\ \citenamefont {Volansky}}]{Essig:2016crl}%
  \BibitemOpen
  \bibfield  {author} {\bibinfo {author} {\bibfnamefont {R.}~\bibnamefont
  {Essig}}, \bibinfo {author} {\bibfnamefont {J.}~\bibnamefont {Mardon}},
  \bibinfo {author} {\bibfnamefont {O.}~\bibnamefont {Slone}}, \ and\ \bibinfo
  {author} {\bibfnamefont {T.}~\bibnamefont {Volansky}},\ }\href {\doibase
  10.1103/PhysRevD.95.056011} {\bibfield  {journal} {\bibinfo  {journal} {Phys.
  Rev.}\ }\textbf {\bibinfo {volume} {D95}},\ \bibinfo {pages} {056011}
  (\bibinfo {year} {2017}{\natexlab{b}})},\ \Eprint
  {http://arxiv.org/abs/1608.02940} {arXiv:1608.02940 [hep-ph]} \BibitemShut
  {NoStop}%
\bibitem [{\citenamefont {Hochberg}\ \emph
  {et~al.}(2017{\natexlab{a}})\citenamefont {Hochberg}, \citenamefont {Kahn},
  \citenamefont {Lisanti}, \citenamefont {Tully},\ and\ \citenamefont
  {Zurek}}]{Hochberg:2016ntt}%
  \BibitemOpen
  \bibfield  {author} {\bibinfo {author} {\bibfnamefont {Y.}~\bibnamefont
  {Hochberg}}, \bibinfo {author} {\bibfnamefont {Y.}~\bibnamefont {Kahn}},
  \bibinfo {author} {\bibfnamefont {M.}~\bibnamefont {Lisanti}}, \bibinfo
  {author} {\bibfnamefont {C.~G.}\ \bibnamefont {Tully}}, \ and\ \bibinfo
  {author} {\bibfnamefont {K.~M.}\ \bibnamefont {Zurek}},\ }\href {\doibase
  10.1016/j.physletb.2017.06.051} {\bibfield  {journal} {\bibinfo  {journal}
  {Phys. Lett.}\ }\textbf {\bibinfo {volume} {B772}},\ \bibinfo {pages} {239}
  (\bibinfo {year} {2017}{\natexlab{a}})},\ \Eprint
  {http://arxiv.org/abs/1606.08849} {arXiv:1606.08849 [hep-ph]} \BibitemShut
  {NoStop}%
\bibitem [{\citenamefont {Hochberg}\ \emph
  {et~al.}(2016{\natexlab{c}})\citenamefont {Hochberg}, \citenamefont {Lin},\
  and\ \citenamefont {Zurek}}]{Hochberg:2016ajh}%
  \BibitemOpen
  \bibfield  {author} {\bibinfo {author} {\bibfnamefont {Y.}~\bibnamefont
  {Hochberg}}, \bibinfo {author} {\bibfnamefont {T.}~\bibnamefont {Lin}}, \
  and\ \bibinfo {author} {\bibfnamefont {K.~M.}\ \bibnamefont {Zurek}},\ }\href
  {\doibase 10.1103/PhysRevD.94.015019} {\bibfield  {journal} {\bibinfo
  {journal} {Phys. Rev.}\ }\textbf {\bibinfo {volume} {D94}},\ \bibinfo {pages}
  {015019} (\bibinfo {year} {2016}{\natexlab{c}})},\ \Eprint
  {http://arxiv.org/abs/1604.06800} {arXiv:1604.06800 [hep-ph]} \BibitemShut
  {NoStop}%
\bibitem [{\citenamefont {Hochberg}\ \emph
  {et~al.}(2017{\natexlab{b}})\citenamefont {Hochberg}, \citenamefont {Lin},\
  and\ \citenamefont {Zurek}}]{Hochberg:2016sqx}%
  \BibitemOpen
  \bibfield  {author} {\bibinfo {author} {\bibfnamefont {Y.}~\bibnamefont
  {Hochberg}}, \bibinfo {author} {\bibfnamefont {T.}~\bibnamefont {Lin}}, \
  and\ \bibinfo {author} {\bibfnamefont {K.~M.}\ \bibnamefont {Zurek}},\ }\href
  {\doibase 10.1103/PhysRevD.95.023013} {\bibfield  {journal} {\bibinfo
  {journal} {Phys. Rev.}\ }\textbf {\bibinfo {volume} {D95}},\ \bibinfo {pages}
  {023013} (\bibinfo {year} {2017}{\natexlab{b}})},\ \Eprint
  {http://arxiv.org/abs/1608.01994} {arXiv:1608.01994 [hep-ph]} \BibitemShut
  {NoStop}%
\bibitem [{\citenamefont {Budnik}\ \emph {et~al.}(2018)\citenamefont {Budnik},
  \citenamefont {Chesnovsky}, \citenamefont {Slone},\ and\ \citenamefont
  {Volansky}}]{Budnik:2017sbu}%
  \BibitemOpen
  \bibfield  {author} {\bibinfo {author} {\bibfnamefont {R.}~\bibnamefont
  {Budnik}}, \bibinfo {author} {\bibfnamefont {O.}~\bibnamefont {Chesnovsky}},
  \bibinfo {author} {\bibfnamefont {O.}~\bibnamefont {Slone}}, \ and\ \bibinfo
  {author} {\bibfnamefont {T.}~\bibnamefont {Volansky}},\ }\href {\doibase
  10.1016/j.physletb.2018.04.063} {\bibfield  {journal} {\bibinfo  {journal}
  {Phys. Lett.}\ }\textbf {\bibinfo {volume} {B782}},\ \bibinfo {pages} {242}
  (\bibinfo {year} {2018})},\ \Eprint {http://arxiv.org/abs/1705.03016}
  {arXiv:1705.03016 [hep-ph]} \BibitemShut {NoStop}%
\bibitem [{\citenamefont {Bunting}\ \emph {et~al.}(2017)\citenamefont
  {Bunting}, \citenamefont {Gratta}, \citenamefont {Melia},\ and\ \citenamefont
  {Rajendran}}]{Bunting:2017net}%
  \BibitemOpen
  \bibfield  {author} {\bibinfo {author} {\bibfnamefont {P.~C.}\ \bibnamefont
  {Bunting}}, \bibinfo {author} {\bibfnamefont {G.}~\bibnamefont {Gratta}},
  \bibinfo {author} {\bibfnamefont {T.}~\bibnamefont {Melia}}, \ and\ \bibinfo
  {author} {\bibfnamefont {S.}~\bibnamefont {Rajendran}},\ }\href {\doibase
  10.1103/PhysRevD.95.095001} {\bibfield  {journal} {\bibinfo  {journal} {Phys.
  Rev.}\ }\textbf {\bibinfo {volume} {D95}},\ \bibinfo {pages} {095001}
  (\bibinfo {year} {2017})},\ \Eprint {http://arxiv.org/abs/1701.06566}
  {arXiv:1701.06566 [hep-ph]} \BibitemShut {NoStop}%
\bibitem [{\citenamefont {Cavoto}\ \emph {et~al.}(2018)\citenamefont {Cavoto},
  \citenamefont {Luchetta},\ and\ \citenamefont {Polosa}}]{Cavoto:2017otc}%
  \BibitemOpen
  \bibfield  {author} {\bibinfo {author} {\bibfnamefont {G.}~\bibnamefont
  {Cavoto}}, \bibinfo {author} {\bibfnamefont {F.}~\bibnamefont {Luchetta}}, \
  and\ \bibinfo {author} {\bibfnamefont {A.~D.}\ \bibnamefont {Polosa}},\
  }\href {\doibase 10.1016/j.physletb.2017.11.064} {\bibfield  {journal}
  {\bibinfo  {journal} {Phys. Lett.}\ }\textbf {\bibinfo {volume} {B776}},\
  \bibinfo {pages} {338} (\bibinfo {year} {2018})},\ \Eprint
  {http://arxiv.org/abs/1706.02487} {arXiv:1706.02487 [hep-ph]} \BibitemShut
  {NoStop}%
\bibitem [{\citenamefont {Fichet}(2018)}]{Fichet:2017bng}%
  \BibitemOpen
  \bibfield  {author} {\bibinfo {author} {\bibfnamefont {S.}~\bibnamefont
  {Fichet}},\ }\href {\doibase 10.1103/PhysRevLett.120.131801} {\bibfield
  {journal} {\bibinfo  {journal} {Phys. Rev. Lett.}\ }\textbf {\bibinfo
  {volume} {120}},\ \bibinfo {pages} {131801} (\bibinfo {year} {2018})},\
  \Eprint {http://arxiv.org/abs/1705.10331} {arXiv:1705.10331 [hep-ph]}
  \BibitemShut {NoStop}%
\bibitem [{\citenamefont {Knapen}\ \emph {et~al.}(2018)\citenamefont {Knapen},
  \citenamefont {Lin}, \citenamefont {Pyle},\ and\ \citenamefont
  {Zurek}}]{Knapen:2017ekk}%
  \BibitemOpen
  \bibfield  {author} {\bibinfo {author} {\bibfnamefont {S.}~\bibnamefont
  {Knapen}}, \bibinfo {author} {\bibfnamefont {T.}~\bibnamefont {Lin}},
  \bibinfo {author} {\bibfnamefont {M.}~\bibnamefont {Pyle}}, \ and\ \bibinfo
  {author} {\bibfnamefont {K.~M.}\ \bibnamefont {Zurek}},\ }\href {\doibase
  10.1016/j.physletb.2018.08.064} {\bibfield  {journal} {\bibinfo  {journal}
  {Phys. Lett.}\ }\textbf {\bibinfo {volume} {B785}},\ \bibinfo {pages} {386}
  (\bibinfo {year} {2018})},\ \Eprint {http://arxiv.org/abs/1712.06598}
  {arXiv:1712.06598 [hep-ph]} \BibitemShut {NoStop}%
\bibitem [{\citenamefont {Hochberg}\ \emph {et~al.}(2018)\citenamefont
  {Hochberg}, \citenamefont {Kahn}, \citenamefont {Lisanti}, \citenamefont
  {Zurek}, \citenamefont {Grushin}, \citenamefont {Ilan}, \citenamefont
  {Griffin}, \citenamefont {Liu}, \citenamefont {Weber},\ and\ \citenamefont
  {Neaton}}]{Hochberg:2017wce}%
  \BibitemOpen
  \bibfield  {author} {\bibinfo {author} {\bibfnamefont {Y.}~\bibnamefont
  {Hochberg}}, \bibinfo {author} {\bibfnamefont {Y.}~\bibnamefont {Kahn}},
  \bibinfo {author} {\bibfnamefont {M.}~\bibnamefont {Lisanti}}, \bibinfo
  {author} {\bibfnamefont {K.~M.}\ \bibnamefont {Zurek}}, \bibinfo {author}
  {\bibfnamefont {A.~G.}\ \bibnamefont {Grushin}}, \bibinfo {author}
  {\bibfnamefont {R.}~\bibnamefont {Ilan}}, \bibinfo {author} {\bibfnamefont
  {S.~M.}\ \bibnamefont {Griffin}}, \bibinfo {author} {\bibfnamefont {Z.-F.}\
  \bibnamefont {Liu}}, \bibinfo {author} {\bibfnamefont {S.~F.}\ \bibnamefont
  {Weber}}, \ and\ \bibinfo {author} {\bibfnamefont {J.~B.}\ \bibnamefont
  {Neaton}},\ }\href {\doibase 10.1103/PhysRevD.97.015004} {\bibfield
  {journal} {\bibinfo  {journal} {Phys. Rev.}\ }\textbf {\bibinfo {volume}
  {D97}},\ \bibinfo {pages} {015004} (\bibinfo {year} {2018})},\ \Eprint
  {http://arxiv.org/abs/1708.08929} {arXiv:1708.08929 [hep-ph]} \BibitemShut
  {NoStop}%
\bibitem [{\citenamefont {Dolan}\ \emph {et~al.}(2018)\citenamefont {Dolan},
  \citenamefont {Kahlhoefer},\ and\ \citenamefont {McCabe}}]{Dolan:2017xbu}%
  \BibitemOpen
  \bibfield  {author} {\bibinfo {author} {\bibfnamefont {M.~J.}\ \bibnamefont
  {Dolan}}, \bibinfo {author} {\bibfnamefont {F.}~\bibnamefont {Kahlhoefer}}, \
  and\ \bibinfo {author} {\bibfnamefont {C.}~\bibnamefont {McCabe}},\ }\href
  {\doibase 10.1103/PhysRevLett.121.101801} {\bibfield  {journal} {\bibinfo
  {journal} {Phys. Rev. Lett.}\ }\textbf {\bibinfo {volume} {121}},\ \bibinfo
  {pages} {101801} (\bibinfo {year} {2018})},\ \Eprint
  {http://arxiv.org/abs/1711.09906} {arXiv:1711.09906 [hep-ph]} \BibitemShut
  {NoStop}%
\bibitem [{\citenamefont {Bringmann}\ and\ \citenamefont
  {Pospelov}(2019)}]{Bringmann:2018cvk}%
  \BibitemOpen
  \bibfield  {author} {\bibinfo {author} {\bibfnamefont {T.}~\bibnamefont
  {Bringmann}}\ and\ \bibinfo {author} {\bibfnamefont {M.}~\bibnamefont
  {Pospelov}},\ }\href {\doibase 10.1103/PhysRevLett.122.171801} {\bibfield
  {journal} {\bibinfo  {journal} {Phys. Rev. Lett.}\ }\textbf {\bibinfo
  {volume} {122}},\ \bibinfo {pages} {171801} (\bibinfo {year} {2019})},\
  \Eprint {http://arxiv.org/abs/1810.10543} {arXiv:1810.10543 [hep-ph]}
  \BibitemShut {NoStop}%
\bibitem [{\citenamefont {Ema}\ \emph {et~al.}(2019)\citenamefont {Ema},
  \citenamefont {Sala},\ and\ \citenamefont {Sato}}]{Ema:2018bih}%
  \BibitemOpen
  \bibfield  {author} {\bibinfo {author} {\bibfnamefont {Y.}~\bibnamefont
  {Ema}}, \bibinfo {author} {\bibfnamefont {F.}~\bibnamefont {Sala}}, \ and\
  \bibinfo {author} {\bibfnamefont {R.}~\bibnamefont {Sato}},\ }\href {\doibase
  10.1103/PhysRevLett.122.181802} {\bibfield  {journal} {\bibinfo  {journal}
  {Phys. Rev. Lett.}\ }\textbf {\bibinfo {volume} {122}},\ \bibinfo {pages}
  {181802} (\bibinfo {year} {2019})},\ \Eprint
  {http://arxiv.org/abs/1811.00520} {arXiv:1811.00520 [hep-ph]} \BibitemShut
  {NoStop}%
\bibitem [{\citenamefont {Emken}\ \emph {et~al.}(2019)\citenamefont {Emken},
  \citenamefont {Essig}, \citenamefont {Kouvaris},\ and\ \citenamefont
  {Sholapurkar}}]{Emken:2019tni}%
  \BibitemOpen
  \bibfield  {author} {\bibinfo {author} {\bibfnamefont {T.}~\bibnamefont
  {Emken}}, \bibinfo {author} {\bibfnamefont {R.}~\bibnamefont {Essig}},
  \bibinfo {author} {\bibfnamefont {C.}~\bibnamefont {Kouvaris}}, \ and\
  \bibinfo {author} {\bibfnamefont {M.}~\bibnamefont {Sholapurkar}},\
  }\href@noop {} {\  (\bibinfo {year} {2019})},\ \Eprint
  {http://arxiv.org/abs/1905.06348} {arXiv:1905.06348 [hep-ph]} \BibitemShut
  {NoStop}%
\bibitem [{\citenamefont {Bell}\ \emph {et~al.}(2019)\citenamefont {Bell},
  \citenamefont {Dent}, \citenamefont {Newstead}, \citenamefont {Sabharwale},\
  and\ \citenamefont {Weiler}}]{Bell:2019egg}%
  \BibitemOpen
  \bibfield  {author} {\bibinfo {author} {\bibfnamefont {N.~F.}\ \bibnamefont
  {Bell}}, \bibinfo {author} {\bibfnamefont {J.~B.}\ \bibnamefont {Dent}},
  \bibinfo {author} {\bibfnamefont {J.~L.}\ \bibnamefont {Newstead}}, \bibinfo
  {author} {\bibfnamefont {S.}~\bibnamefont {Sabharwale}}, \ and\ \bibinfo
  {author} {\bibfnamefont {T.~J.}\ \bibnamefont {Weiler}},\ }\href@noop {} {\
  (\bibinfo {year} {2019})},\ \Eprint {http://arxiv.org/abs/1905.00046}
  {arXiv:1905.00046 [hep-ph]} \BibitemShut {NoStop}%
\bibitem [{\citenamefont {Cappiello}\ and\ \citenamefont
  {Beacom}(2019)}]{Cappiello:2019qsw}%
  \BibitemOpen
  \bibfield  {author} {\bibinfo {author} {\bibfnamefont {C.}~\bibnamefont
  {Cappiello}}\ and\ \bibinfo {author} {\bibfnamefont {J.~F.}\ \bibnamefont
  {Beacom}},\ }\href@noop {} {\  (\bibinfo {year} {2019})},\ \Eprint
  {http://arxiv.org/abs/1906.11283} {arXiv:1906.11283 [hep-ph]} \BibitemShut
  {NoStop}%
\bibitem [{\citenamefont {Essig}\ \emph {et~al.}()\citenamefont {Essig},
  \citenamefont {Matzi}, \citenamefont {Pradler}, \citenamefont {Sholapurkar},\
  and\ \citenamefont {Yu}}]{MigdalPRD}%
  \BibitemOpen
  \bibfield  {author} {\bibinfo {author} {\bibfnamefont {R.}~\bibnamefont
  {Essig}}, \bibinfo {author} {\bibfnamefont {L.}~\bibnamefont {Matzi}},
  \bibinfo {author} {\bibfnamefont {J.}~\bibnamefont {Pradler}}, \bibinfo
  {author} {\bibfnamefont {M.}~\bibnamefont {Sholapurkar}}, \ and\ \bibinfo
  {author} {\bibfnamefont {T.-T.}\ \bibnamefont {Yu}},\ }\href@noop {} {\
  }\bibinfo {note} {In preparation}\BibitemShut {NoStop}%
\bibitem [{\citenamefont {Gu}(2008)}]{doi:10.1139/p07-197}%
  \BibitemOpen
  \bibfield  {author} {\bibinfo {author} {\bibfnamefont {M.~F.}\ \bibnamefont
  {Gu}},\ }\href {\doibase 10.1139/p07-197} {\bibfield  {journal} {\bibinfo
  {journal} {Canadian Journal of Physics}\ }\textbf {\bibinfo {volume} {86}},\
  \bibinfo {pages} {675} (\bibinfo {year} {2008})},\ \Eprint
  {http://arxiv.org/abs/https://doi.org/10.1139/p07-197}
  {https://doi.org/10.1139/p07-197} \BibitemShut {NoStop}%
\bibitem [{\citenamefont {Roberts}\ \emph {et~al.}(2016)\citenamefont
  {Roberts}, \citenamefont {Dzuba}, \citenamefont {Flambaum}, \citenamefont
  {Pospelov},\ and\ \citenamefont {Stadnik}}]{Roberts:2016xfw}%
  \BibitemOpen
  \bibfield  {author} {\bibinfo {author} {\bibfnamefont {B.~M.}\ \bibnamefont
  {Roberts}}, \bibinfo {author} {\bibfnamefont {V.~A.}\ \bibnamefont {Dzuba}},
  \bibinfo {author} {\bibfnamefont {V.~V.}\ \bibnamefont {Flambaum}}, \bibinfo
  {author} {\bibfnamefont {M.}~\bibnamefont {Pospelov}}, \ and\ \bibinfo
  {author} {\bibfnamefont {Y.~V.}\ \bibnamefont {Stadnik}},\ }\href {\doibase
  10.1103/PhysRevD.93.115037} {\bibfield  {journal} {\bibinfo  {journal} {Phys.
  Rev.}\ }\textbf {\bibinfo {volume} {D93}},\ \bibinfo {pages} {115037}
  (\bibinfo {year} {2016})},\ \Eprint {http://arxiv.org/abs/1604.04559}
  {arXiv:1604.04559 [hep-ph]} \BibitemShut {NoStop}%
\bibitem [{\citenamefont {Lewin}\ and\ \citenamefont
  {Smith}(1996)}]{Lewin:1995rx}%
  \BibitemOpen
  \bibfield  {author} {\bibinfo {author} {\bibfnamefont {J.~D.}\ \bibnamefont
  {Lewin}}\ and\ \bibinfo {author} {\bibfnamefont {P.~F.}\ \bibnamefont
  {Smith}},\ }\href {\doibase 10.1016/S0927-6505(96)00047-3} {\bibfield
  {journal} {\bibinfo  {journal} {Astropart. Phys.}\ }\textbf {\bibinfo
  {volume} {6}},\ \bibinfo {pages} {87} (\bibinfo {year} {1996})}\BibitemShut
  {NoStop}%
\bibitem [{\citenamefont {Abdelhameed}\ \emph {et~al.}(2019)\citenamefont
  {Abdelhameed} \emph {et~al.}}]{Abdelhameed:2019hmk}%
  \BibitemOpen
  \bibfield  {author} {\bibinfo {author} {\bibfnamefont {A.~H.}\ \bibnamefont
  {Abdelhameed}} \emph {et~al.} (\bibinfo {collaboration} {CRESST}),\
  }\href@noop {} {\  (\bibinfo {year} {2019})},\ \Eprint
  {http://arxiv.org/abs/1904.00498} {arXiv:1904.00498 [astro-ph.CO]}
  \BibitemShut {NoStop}%
\bibitem [{\citenamefont {Collar}(2018)}]{Collar:2018ydf}%
  \BibitemOpen
  \bibfield  {author} {\bibinfo {author} {\bibfnamefont {J.~I.}\ \bibnamefont
  {Collar}},\ }\href {\doibase 10.1103/PhysRevD.98.023005,
  10.1103/PHYSREVD.98.023005} {\bibfield  {journal} {\bibinfo  {journal} {Phys.
  Rev.}\ }\textbf {\bibinfo {volume} {D98}},\ \bibinfo {pages} {023005}
  (\bibinfo {year} {2018})},\ \Eprint {http://arxiv.org/abs/1805.02646}
  {arXiv:1805.02646 [astro-ph.CO]} \BibitemShut {NoStop}%
\bibitem [{\citenamefont {Agnes}\ \emph
  {et~al.}(2018{\natexlab{b}})\citenamefont {Agnes} \emph
  {et~al.}}]{Agnes:2018ves}%
  \BibitemOpen
  \bibfield  {author} {\bibinfo {author} {\bibfnamefont {P.}~\bibnamefont
  {Agnes}} \emph {et~al.} (\bibinfo {collaboration} {DarkSide}),\ }\href
  {\doibase 10.1103/PhysRevLett.121.081307} {\bibfield  {journal} {\bibinfo
  {journal} {Phys. Rev. Lett.}\ }\textbf {\bibinfo {volume} {121}},\ \bibinfo
  {pages} {081307} (\bibinfo {year} {2018}{\natexlab{b}})},\ \Eprint
  {http://arxiv.org/abs/1802.06994} {arXiv:1802.06994 [astro-ph.HE]}
  \BibitemShut {NoStop}%
\bibitem [{\citenamefont {Aprile}\ \emph {et~al.}(2018)\citenamefont {Aprile}
  \emph {et~al.}}]{Aprile:2018dbl}%
  \BibitemOpen
  \bibfield  {author} {\bibinfo {author} {\bibfnamefont {E.}~\bibnamefont
  {Aprile}} \emph {et~al.} (\bibinfo {collaboration} {XENON}),\ }\href
  {\doibase 10.1103/PhysRevLett.121.111302} {\bibfield  {journal} {\bibinfo
  {journal} {Phys. Rev. Lett.}\ }\textbf {\bibinfo {volume} {121}},\ \bibinfo
  {pages} {111302} (\bibinfo {year} {2018})},\ \Eprint
  {http://arxiv.org/abs/1805.12562} {arXiv:1805.12562 [astro-ph.CO]}
  \BibitemShut {NoStop}%
\bibitem [{\citenamefont {Aprile}\ \emph
  {et~al.}(2019{\natexlab{b}})\citenamefont {Aprile} \emph
  {et~al.}}]{Aprile:2019xxb}%
  \BibitemOpen
  \bibfield  {author} {\bibinfo {author} {\bibfnamefont {E.}~\bibnamefont
  {Aprile}} \emph {et~al.},\ }\href@noop {} {\  (\bibinfo {year}
  {2019}{\natexlab{b}})},\ \Eprint {http://arxiv.org/abs/1907.11485}
  {arXiv:1907.11485 [hep-ex]} \BibitemShut {NoStop}%
\bibitem [{\citenamefont {Ren}\ \emph {et~al.}(2018)\citenamefont {Ren} \emph
  {et~al.}}]{Ren:2018gyx}%
  \BibitemOpen
  \bibfield  {author} {\bibinfo {author} {\bibfnamefont {X.}~\bibnamefont
  {Ren}} \emph {et~al.} (\bibinfo {collaboration} {PandaX-II}),\ }\href
  {\doibase 10.1103/PhysRevLett.121.021304} {\bibfield  {journal} {\bibinfo
  {journal} {Phys. Rev. Lett.}\ }\textbf {\bibinfo {volume} {121}},\ \bibinfo
  {pages} {021304} (\bibinfo {year} {2018})},\ \Eprint
  {http://arxiv.org/abs/1802.06912} {arXiv:1802.06912 [hep-ph]} \BibitemShut
  {NoStop}%
\bibitem [{QEd()}]{QEdark}%
  \BibitemOpen
  \href@noop {} {\enquote {\bibinfo {title} {{\tt{QEDark}}},}\ }\bibinfo
  {howpublished} {\url{http://ddldm.physics.sunysb.edu/}}\BibitemShut {NoStop}%
\bibitem [{\citenamefont {Dahl}(2009)}]{Dahl:2009nta}%
  \BibitemOpen
  \bibfield  {author} {\bibinfo {author} {\bibfnamefont {C.~E.}\ \bibnamefont
  {Dahl}},\ }\emph {\bibinfo {title} {{The physics of background discrimination
  in liquid xenon, and first results from Xenon10 in the hunt for WIMP dark
  matter}}},\ \href
  {https://www.princeton.edu/physics/graduate-program/theses/theses-from-2009/E.Dahlthesis.pdf}
  {Ph.D. thesis},\ \bibinfo  {school} {Princeton U.} (\bibinfo {year}
  {2009})\BibitemShut {NoStop}%
\bibitem [{\citenamefont {Akerib}\ \emph
  {et~al.}(2017{\natexlab{a}})\citenamefont {Akerib} \emph
  {et~al.}}]{Akerib:2016qlr}%
  \BibitemOpen
  \bibfield  {author} {\bibinfo {author} {\bibfnamefont {D.~S.}\ \bibnamefont
  {Akerib}} \emph {et~al.} (\bibinfo {collaboration} {LUX}),\ }\href {\doibase
  10.1103/PhysRevD.95.012008} {\bibfield  {journal} {\bibinfo  {journal} {Phys.
  Rev.}\ }\textbf {\bibinfo {volume} {D95}},\ \bibinfo {pages} {012008}
  (\bibinfo {year} {2017}{\natexlab{a}})},\ \Eprint
  {http://arxiv.org/abs/1610.02076} {arXiv:1610.02076 [physics.ins-det]}
  \BibitemShut {NoStop}%
\bibitem [{\citenamefont {Akerib}\ \emph
  {et~al.}(2017{\natexlab{b}})\citenamefont {Akerib} \emph
  {et~al.}}]{Akerib:2017hph}%
  \BibitemOpen
  \bibfield  {author} {\bibinfo {author} {\bibfnamefont {D.~S.}\ \bibnamefont
  {Akerib}} \emph {et~al.} (\bibinfo {collaboration} {LUX}),\ }\href {\doibase
  10.1103/PhysRevD.96.112011} {\bibfield  {journal} {\bibinfo  {journal} {Phys.
  Rev.}\ }\textbf {\bibinfo {volume} {D96}},\ \bibinfo {pages} {112011}
  (\bibinfo {year} {2017}{\natexlab{b}})},\ \Eprint
  {http://arxiv.org/abs/1709.00800} {arXiv:1709.00800 [physics.ins-det]}
  \BibitemShut {NoStop}%
\bibitem [{\citenamefont {Aprile}\ \emph
  {et~al.}(2014{\natexlab{b}})\citenamefont {Aprile} \emph
  {et~al.}}]{Aprile:2013blg}%
  \BibitemOpen
  \bibfield  {author} {\bibinfo {author} {\bibfnamefont {E.}~\bibnamefont
  {Aprile}} \emph {et~al.} (\bibinfo {collaboration} {XENON100}),\ }\href
  {\doibase 10.1088/0954-3899/41/3/035201} {\bibfield  {journal} {\bibinfo
  {journal} {J. Phys.}\ }\textbf {\bibinfo {volume} {G41}},\ \bibinfo {pages}
  {035201} (\bibinfo {year} {2014}{\natexlab{b}})},\ \Eprint
  {http://arxiv.org/abs/1311.1088} {arXiv:1311.1088 [physics.ins-det]}
  \BibitemShut {NoStop}%
\bibitem [{\citenamefont {Angle}\ \emph {et~al.}(2011)\citenamefont {Angle}
  \emph {et~al.}}]{Angle:2011th}%
  \BibitemOpen
  \bibfield  {author} {\bibinfo {author} {\bibfnamefont {J.}~\bibnamefont
  {Angle}} \emph {et~al.} (\bibinfo {collaboration} {XENON10}),\ }\href
  {\doibase 10.1103/PhysRevLett.110.249901, 10.1103/PhysRevLett.107.051301}
  {\bibfield  {journal} {\bibinfo  {journal} {Phys. Rev. Lett.}\ }\textbf
  {\bibinfo {volume} {107}},\ \bibinfo {pages} {051301} (\bibinfo {year}
  {2011})},\ \bibinfo {note} {[Erratum: Phys. Rev. Lett.110,249901(2013)]},\
  \Eprint {http://arxiv.org/abs/1104.3088} {arXiv:1104.3088 [astro-ph.CO]}
  \BibitemShut {NoStop}%
\bibitem [{\citenamefont {Aprile}\ \emph {et~al.}(2016)\citenamefont {Aprile}
  \emph {et~al.}}]{Aprile:2016wwo}%
  \BibitemOpen
  \bibfield  {author} {\bibinfo {author} {\bibfnamefont {E.}~\bibnamefont
  {Aprile}} \emph {et~al.} (\bibinfo {collaboration} {XENON}),\ }\href
  {\doibase 10.1103/PhysRevD.94.092001, 10.1103/PhysRevD.95.059901} {\bibfield
  {journal} {\bibinfo  {journal} {Phys. Rev.}\ }\textbf {\bibinfo {volume}
  {D94}},\ \bibinfo {pages} {092001} (\bibinfo {year} {2016})},\ \bibinfo
  {note} {[Erratum: Phys. Rev.D95,no.5,059901(2017)]},\ \Eprint
  {http://arxiv.org/abs/1605.06262} {arXiv:1605.06262 [astro-ph.CO]}
  \BibitemShut {NoStop}%
\bibitem [{\citenamefont {Essig}\ \emph {et~al.}(2018)\citenamefont {Essig},
  \citenamefont {Sholapurkar},\ and\ \citenamefont {Yu}}]{Essig:2018tss}%
  \BibitemOpen
  \bibfield  {author} {\bibinfo {author} {\bibfnamefont {R.}~\bibnamefont
  {Essig}}, \bibinfo {author} {\bibfnamefont {M.}~\bibnamefont {Sholapurkar}},
  \ and\ \bibinfo {author} {\bibfnamefont {T.-T.}\ \bibnamefont {Yu}},\ }\href
  {\doibase 10.1103/PhysRevD.97.095029} {\bibfield  {journal} {\bibinfo
  {journal} {Phys. Rev.}\ }\textbf {\bibinfo {volume} {D97}},\ \bibinfo {pages}
  {095029} (\bibinfo {year} {2018})},\ \Eprint
  {http://arxiv.org/abs/1801.10159} {arXiv:1801.10159 [hep-ph]} \BibitemShut
  {NoStop}%
\bibitem [{\citenamefont {Battaglieri}\ \emph {et~al.}(2017)\citenamefont
  {Battaglieri} \emph {et~al.}}]{Battaglieri:2017aum}%
  \BibitemOpen
  \bibfield  {author} {\bibinfo {author} {\bibfnamefont {M.}~\bibnamefont
  {Battaglieri}} \emph {et~al.},\ }in\ \href
  {http://lss.fnal.gov/archive/2017/conf/fermilab-conf-17-282-ae-ppd-t.pdf}
  {\emph {\bibinfo {booktitle} {{U.S. Cosmic Visions: New Ideas in Dark Matter
  College Park, MD, USA, March 23-25, 2017}}}}\ (\bibinfo {year} {2017})\
  \Eprint {http://arxiv.org/abs/1707.04591} {arXiv:1707.04591 [hep-ph]}
  \BibitemShut {NoStop}%
\bibitem [{\citenamefont {Baxter}\ \emph {et~al.}(2019)\citenamefont {Baxter},
  \citenamefont {Kahn},\ and\ \citenamefont {Krnjaic}}]{Baxter:2019pnz}%
  \BibitemOpen
  \bibfield  {author} {\bibinfo {author} {\bibfnamefont {D.}~\bibnamefont
  {Baxter}}, \bibinfo {author} {\bibfnamefont {Y.}~\bibnamefont {Kahn}}, \ and\
  \bibinfo {author} {\bibfnamefont {G.}~\bibnamefont {Krnjaic}},\ }\href@noop
  {} {\  (\bibinfo {year} {2019})},\ \Eprint {http://arxiv.org/abs/1908.00012}
  {arXiv:1908.00012 [hep-ph]} \BibitemShut {NoStop}%
\end{thebibliography}%

\end{document}